\documentclass[prd,twocolumn,showpacs,preprintnumbers,superscriptaddress,floatfix,nofootinbib]{revtex4-2}

\usepackage[left]{lineno}

\usepackage{amsfonts}
\usepackage{multirow}
\usepackage{subfigure}

\usepackage{graphicx,,booktabs}
\usepackage{amssymb}
\usepackage{amsmath,txfonts,ulem}
\usepackage{graphicx}
\usepackage{dcolumn}
\usepackage{color}
\usepackage{bm}
\usepackage{ulem}

\usepackage{enumerate}

\usepackage{appendix}
\usepackage[colorlinks,
 citecolor=blue,
 anchorcolor=green,
 menucolor=orange,
 linkcolor=blue,
 filecolor=red,
 runcolor=pink,
 urlcolor=blue,
 frenchlinks=red]{hyperref} 

\allowdisplaybreaks[4]

\begin{document}

\title{\boldmath   Suppressing Coherent Synchrotron Radiation Effects in Chicane Bunch Compressors } %with suppressed   coherent synchrotron radiation  effects %}

\author{ Fancong Zeng}
%\email{ zengfc@ihep.ac.cn}
\affiliation{ Key Laboratory of Particle Acceleration Physics and Technology, Institute of High Energy Physics, Chinese Academy of Sciences, and University of Chinese Academy of Sciences, Beijing 100049, China } 
  
\author{Yi Jiao}
\email{jiaoyi@ihep.ac.cn}
\affiliation{ Key Laboratory of Particle Acceleration Physics and Technology, Institute of High Energy Physics, Chinese Academy of Sciences, and University of Chinese Academy of Sciences, Beijing 100049, China }

\author{Weihang Liu}
\affiliation{China Spallation Neutron Source, Institute of High Energy Physics,
Chinese Academy of Sciences, Dongguan 523803, China }

\author{Cheng-Ying Tsai}
\affiliation{ School of Electrical and Electronic Engineering,
Huazhong University of Science and Technology, Wuhan 430074, China }

\date{\today}
\begin{abstract} 
The most significant advances in the accelerator-based light sources (i.e., x-ray free electron lasers) are driven by the production of the high final peak current in the last several decades. As a prerequisite to attain the proposed high brightness, the symmetric C-chicane bunch compressor is typically exploited due to its simplicity, efficiency, and natural  dispersion-free feature at all orders. However, during bunch compression for a high peak current requirement, a main contributing factor to the transverse emittance degradation is the emission of the coherent synchrotron radiation (CSR). Suppressing this effect is necessary to preserve the beam phase-space quality.   To this end, this paper presents an analysis of one-dimensional CSR point-kick and derives the cancellation conditions in terms of   compression factor. 
The CSR cancelation conditions indicate an asymmetric geometric design. We demonstrate concrete schemes for asymmetric C- and S-chicanes, and verify the CSR cancelation conditions using integration methods and ELEGANT simulations.        
 Furthermore,  the proposed asymmetric C- and S-chicanes  can drastically suppress the emittance growth  compared with the symmetric ones  with identical bunch compression goals.  %, up to  one order of magnitude.   
 
\end{abstract}
 
\maketitle

 \section{Introduction}
  The advent of accelerators and accelerator-based  light sources  has revolutionised the   development of science, industry, medicine, and materials research.    
In contemporary accelerators, electron bunch compressors play a pivotal role, with wide applications in linear colliders \cite{ILC}, linacs \cite{SLAC}, beam-driven plasma-wakefield accelerators \cite{Esarey:2009zz,Lindstrom:2020pzp}, and    significant application in free electron lasers  (FELs)    
 \cite{Borland:2001xv,Three-chicanes,Swiss-report,Spring-report,LCLS-report}.  
  Combined with the position-energy correlation provided by the RF cavity, the following dispersive element helps to convert the energy difference into a difference in the time of flight of the particles. This causes the   particles at the head and tail to become closer, enabling beam compression.
 
Currently,  the symmetric C-chicanes with four bending magnets  are the most commonly used bunch compressors in accelerator systems, because such chicanes are simple, effective and of course, dispersion-free at all orders   \cite{Raubenheimer}. 
However, with the high compression factor and $\sim$kiloampere level peak current required for FELs, the expected FEL performance is hard to preserve due to the coherent synchrotron radiation (CSR) effect \cite{Saldin:1996gs,Dohlus:1996wr,cite:Heung,Heifets:2002qt,Braun:2000dv}.
 When an electron bunch travels in a curved path, it is possible for the coherent radiation emitted by the trailing particles can interact with the leading particles within the dipole, spoiling the transverse beam quality. This effect is particularly noticeable in cases of high bunch current and short bunch length.  
As a consequence of the CSR effect,
the emission of the CSR  leads to transverse emittance growth, which is more obvious in the final two dipoles of the chicane, where the bunch becomes shorter  \cite{Borland:2001xv}. 
Not only are there transverse effects, but also the short-range wake driven microbunching instability (MBI) resulting from density-energy modulations, in the longitudinal plane,  can also be a potentially  detrimental phenomenon  \cite{Tsai:2017hef,Roussel:2015vix,Tsai:2020ddc,Tsai:2017wyq,Heifets:2002qt,Huang:2002kp,Venturini:2007zzd,Venturini:2007zzb,caity}.  
  To alleviate this difficulty,  various efforts have been stimulated, including analytical, numerical, and experimental studies  to suppress the CSR effect in the past decade \cite{PRL,Mitchell:2013tla,Penco:2014sza,Cornacchia:2002xb,Xiang:2011zzd,Stulle:2007se,Venturini:2016dqu,DiMitri:2013qj,Jiao:2014gja,Hajima:2003,Emma:1997hj,Jing:2013cma,Dohlus:1998,Zhang:2023cgl,Mitri-DBA,DiMitri:2013qj,DiMitri:2016gia,Khan:2022tkg,Hajima,Huang:2014bja}. 

To suppress the CSR-induced emittance growth, one mitigation approach exploited the correlation between the CSR and the longitudinal distribution of the beam, using longitudinal shaping   \cite{Penco:2014sza,Mitchell:2013tla} or using a longitudinal transverse emittance  exchanger  \cite{Cornacchia:2002xb,Xiang:2011zzd},  or other complex means.  Experimentally, Ref. \cite{PRL}   observed   that CSR wakes  in a bending magnet could be minimised by using shielding plates. 
On the other hand, 
 the approach of suppressing CSR by manipulating beam optics has sparked continuing research interest.  The optical balance method was first proposed by  D. Douglas \cite{Dohlus:1998}, and further developed by the Courant-Snyder formalism analysis  \cite{DiMitri:2013qj}.  
Moreover,  emittance dilution can be compensated for in a single achromatic cell by matching the beam envelope to the CSR-induced dispersion   \cite{Hajima}.   
 An investigation  proved  that the beam envelope matching and  Courant-Snyder  analysis are equivalent to each other, and indicated  that the complicated CSR in dipole can be evaluated with  a $(x,x^{\prime})$ 2D point-kick analysis \cite{Jiao:2014gja}. 
The optical balance method  and further study have been  successfully applied to many scenarios, e.g.,   spreader, double-bend achromats (DBAs), triple-bend achromats (TBAs)  \cite{Huang:2014bja,Mitri-DBA,DiMitri:2013qj}, 
 and even compression systems,  e.g., DBA-based compressors \cite{Zhang:2023cgl}.

The attempts of cancelling the CSR-driven emittance excitation in chicanes have been a long-standing question. 
%This is because the chicane bunch compressor  is  commonly used, especially in linacs.  
Classified from the four-bend chicane geometries, there are four  chicane designs available, including  symmetric (asymmetric)  C- and S-chicanes. Additionally, if the number of dipoles is not limited,  five-    and   six-bend  chicanes  are also proposed or applied  \cite{Antipov:2021eko,Khan:2022tkg,Stulle:2007se,flash:2007}.
A symmetric chicane (including symmetric four- and six-bend chicanes) is easy to design for a fixed design goal, whereas almost all asymmetric chicanes are given by scanning chicane parameters \cite{Khan:2022tkg,Stulle:2007se}.  
  For example,  
one study  \cite{Khan:2022tkg}   demonstrated a  chicane  with five dipoles of equal length and different bend angles, to  suppress the CSR-induced emittance degradation.   Compared with the symmetric C-chicane, the five-bend design achieves much better emittance suppression, suggesting the potential of increasing the degrees of freedom for a chicane to suppress the CSR effect.    
The parameter scanning method may be effective for a specific chicane design.  Here we study the CSR-inmuned chicanes on a consolidated basis of theoretical guidance, making our results easily transferable to arbitrary chicane design. 

Inspired by the successful application of the point-kick model to a DBA-based compressor \cite{Zhang:2023cgl}, this paper uses the same model to calculate the CSR effect for a chicane bunch compressor.
However,  due to the accumulation of the CSR effect at the chicane exit and  the bunch length variation during compression, the theoretical design of a CSR-immune chicane is much more difficult.  In this respect,  the point-kick model \cite{Jiao:2014gja} can greatly simplify the CSR effect and  compression process.  This allows us to  derive analytical CSR cancellation conditions for the chicane bunch compressor with the aid of the point-kick model. %, which is  commonly used component,  especially in linacs.  

%The purpose of this paper is to conduct a systematic study for four-bend chicane bunch compressors, and based on a more consolidated  foundation of theoretical guidance, enabling our result   easily transferable to arbitrary chicane designs.  
 %Our design retains the beneficial feature that a chicane constructed from four dipoles intrinsically cancels the dispersion.  
 The purpose of this paper is to conduct a four-bend chicane, which consists of four dipoles separated by three drifts. It is assumed that the four bending angles and three drift lengths are free variables.  
Indeed, these variables are not completely free, as they must satisfy the achromatic condition, CSR cancellation conditions, and ensure that the beam returns to its horizontal trajectory (called  ``beam collinear condition").
 
  %, albeit with necessary assumptions.  
%Note that these conditions are critical and will dominate the primary design of the  chicanes. 

For the sake of simplicity, we adopt four assumptions. We refer to the  assumptions in the  theoretical analysis in \cite{Zhang:2023cgl} for a review. 
  First, we assume an electron bunch of   Gaussian temporal distribution. %where Eq. \ref{eq:kk} applied. 
Second, we assume a linear compression process during the electron bunch passing through the RF cavity and the chicane. 
Third,   the  analysis of the one-dimensional CSR  physical   model  is restricted to the steady-state regime and in free space without beam pipe shielding, and  for the moment excludes other effects such as transient CSR and space charge effects.  Lastly, we  ignore the influence of the conducting walls of the vacuum chamber. %Nevertheless, this effect can be accounted for with the method shown in  \cite{Stulle:2007se}. 
In this paper, we do not explore these assumptions any further and concentrate exclusively  on the CSR effect in   bunch compression. 
 
 This paper is organized as follows.  With the CSR point-kick analysis, the conditions for suppressing the emittance growth due to CSR in a chicane can be obtained, as introduced in Sec. \ref{section2}. After  satisfying the achromatic condition, CSR cancellation conditions, and beam collinear condition, we find that the chicane has an excessive number of degrees of freedom.
 To specify the design of the chicane,     in Sec. \ref{C-},  
we   use  two identical dipoles to mimic the structure of a symmetric C-chicane.
The calculation result shows that  a  ``negative drift'' between the 2nd and 3rd dipoles is required. % to cancel the CSR net point kick.  
Then the question arises, whether there exists a chicane that can achieve  the CSR cancelation with all positive drifts between the dipoles. 
 To this end,  we demonstrate that it is a S-chicane geometry, which has never been proved analytically   (Sec. \ref{sub:S}).  The  specific asymmetric S-chicane design are introduced in Sec. \ref{S-}. 
%For practical consideration,..... the bending angle condition in Sec. \ref{C-} is replaced by a position restriction, while the other condition remains unchanged in Sec. \ref{S-}. 
In addition, both the integration method and   ELEGANT particle tracking  are used to verify our calculation results. We also compare the emittance growth of the symmetric C- and S-chicanes with that of the asymmetric ones, while maintaining identical bunch compression targets in Sec. \ref{sec.com}. %The results of our analytical studies  agree with all the verifications basically. 
Summary and dicussion are presented in Sec. \ref{summary}.  

%To provide constraints on the chicane design,  several conditions are added, as introduced in  Secs.  \ref{C-} and \ref{S-}.  

%numerical  verifications for the asymmetric $C$- and $S$-chicane are carried out using an integration method and  ELEGANT  particle tracking. 
%In the remainder of Sec.    \ref{S-}, 
%we compare the   emittance growth between   the symmetric $C$- and $S$-chicane and the asymmetric ones with identical bunch compression targets. 
% Studies show a good efficiency in suppressing the emittance growth. The results from our analytical studies demonstrate good agreement with the all verifications.

\section{ FORMALISM} \label{section2}

\subsection{chicane optics}\label{subsec2.1}
	First, an overview of  the chicane linear optics is given. 
We consider a chicane consisting of four different dipoles separated by  three  adjustable drifts. The total transfer matrix of such chicane can be written as
\begin{equation}\label{eq:mtot}
	M_{tot}= R_{B4} R_{d3} R_{B3} R_{d2} R_{B2} R_{d1} R_{B1}~.
\end{equation}
The $4\times 4$ transfer matrix of a small-angle  dipole
  $R_{Bi}$   ($i=1,2,3,4$)  in the  horizontal and longitudinal planes ($x,x^{\prime}, z,\delta)$ can be expressed by
\begin{equation}\label{eq:RB}
	R_{Bi}=	\left[\begin{array}{cccc}
		1 & \rho_i \theta_i & 0 &  \frac{ \rho_i \theta_i^2  }{2} \\
		0 & 1 & 0 & \theta_i \\
		\theta_i & \frac{ \rho_i \theta_i^2}{2}  & 1 &  \frac{ \rho_i   \theta_i^3}{6}  \\
		0 & 0 & 0 & 1
	\end{array}\right],
\end{equation}
where   $\rho_i  $ and $\theta_i$ are the  beam bending radius and  angle, respectively. %(while keeping $L_{B}=\rho \theta$ constant).  
And the transfer matrix of a half dipole  can be obtained by replacing $\theta_i$ with $\theta_i/2$.
We write the transfer matrix $R_{di}$ of the drift as
\begin{equation}\label{eq:Ld}
	R_{di}=	\left[\begin{array}{cccc}
		1 &L_{di} & 0 &  0 \\
		0  & 1  & 0 & 0 \\
		0 & 0 &1 & 0  \\
		0 & 0 & 0 & 1
	\end{array}\right],
\end{equation}
where $L_{di}$ ($i=1,2,3 $)  is the drift  length.  

We   aim to study the achromatic condition and the beam collinear condition uncovered by the chicane optics. With the total transfer matrix  in Eq.  \eqref{eq:mtot}, the achromatic condition $(M_{tot})_{16}=0$, $(M_{tot})_{26}=0$ and  the first-order  momentum compaction   $R_{56}^{s_0\to s_f} = (M_{tot})_{56}$ for the chicane can   be calculated with
\begin{equation}\label{eq:achromatic-condition-all} 
	\begin{aligned} 
(M_{tot})_{16}=	& L_{d1}\theta_1+L_{d2}(\theta_1+\theta_2)
+L_{d3}(\theta_1+\theta_2+\theta_3)\\
+&\frac{\theta_4}{2} L_{B4}+\frac{\theta_3}{2}(   L_{B3}+2L_{B4})+   \frac{\theta_2}{2}(  L_{B2}+2L_{B3}+2L_{B4})  \\
+&\frac{\theta_1}{2}(L_{B1}+2L_{B2}+2L_{B3}+2L_{B4})=0, \\
	(M_{tot})_{26}=&\theta_1+\theta_2+\theta_3+\theta_4=0,\\
 	R_{56}^{s_0\to s_f} =	&  L_{d1} \theta_1 \theta_2+L_{d3} \theta_3 \theta_4 +\frac{L_{B1} }{6}   \theta_1(3 \theta_2+ \theta_1 ) \\
+&\frac{L_{B2} }{6}   \theta_2( 3\theta_1+  \theta_2) +\frac{L_{B3} }{6}   \theta_3( 3\theta_4+  \theta_3) +\frac{L_{B4} }{6}   \theta_4 ( 3\theta_3+  \theta_4),
		\end{aligned}
\end{equation}
where $L_{Bi}=\rho_i \theta_i$ is the dipole length, and the  chicane entrance and exit are denoted by subscripts ``$s_0$" and ``$s_f$"  respectively. % \textcolor{red}{The compression level for a chicane is mainly reflected in the value of    $ 	R_{56}^{s_0\to s_f}  $. }  
The first two members of Eq. \eqref{eq:achromatic-condition-all}   have a double meaning. 
As a key feature of a lattice with only dipoles, the beam collinear condition  (or called   ``geometric horizontal condition")  that the horizontal trajectory of the beam is coaxial before and after passing through the chicane, coincides  with the achromatic condition.  
 %the trajectory after leaving the chicane is coaxial with the trajectory at the entrance, coincides  with the achromatic condition. 
For this reason, the beam collinear condition can be  achieved naturally.
For the length of bending magnet much shorter than the drift length, the dipole length $L_{Bi}$ is typically neglected, which is appropriate for  most chicane design scenarios. Thus Eq. \eqref{eq:achromatic-condition-all} can lead to simpler expressions as 
\begin{equation}\label{eq:achromatic-condition} 
	\begin{aligned} 
	(M_{tot})_{16}=&L_{d1}\theta_1+L_{d2}(\theta_1+\theta_2)+L_{d3}(\theta_1+\theta_2+\theta_3) =0,\\
	(M_{tot})_{26}=	& \theta_1+\theta_2+\theta_3+\theta_4 =0,\\
	R_{56}^{s_0\to s_f} =	&  L_{d1} \theta_1 \theta_2+L_{d3} \theta_3 \theta_4. 
		\end{aligned}
\end{equation}  
  
Dispersion cancellation is essential for chicane design to prevent dispersion-induced emittance growth caused by  the momentum dispersive effect. 
In the following, we focus more on  the CSR cancelation conditions for a chicane bunch compressor. 
%the CSR-induced emittance amplification for a chicane bunch compressor. 

\subsection{ Application of CSR point-kick model to chicane     }\label{sub:kick}
\begin{figure}[h]
	\centering
	\includegraphics[scale=0.22]{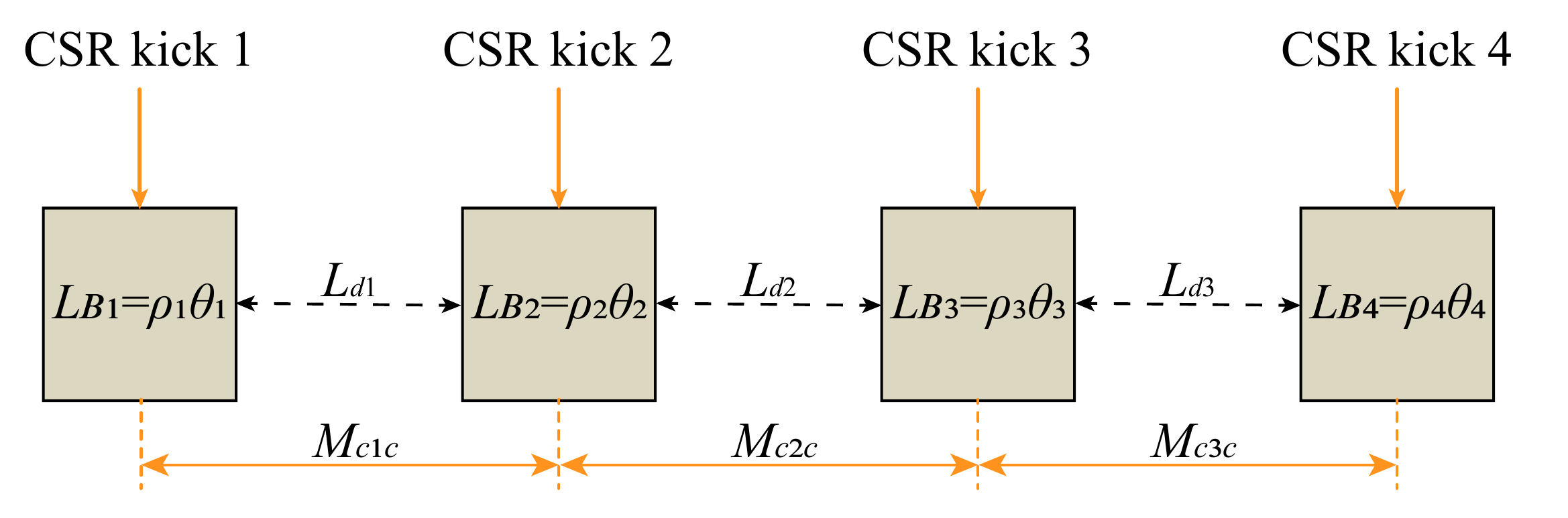}
	\caption{ Schematic of a  chicane for the two-dimensional point-kick analysis of the	CSR effect. The arrows point to  the centers of the   dipoles.   $M_{c1c}, M_{c2c} $ and $M_{c3c}$  represent the transfer matrix from the center of 1st dipole   to 2nd dipole, the center of  2nd dipole   to 3rd dipole, and the center of 3rd dipole to 4th dipole, respectively. }   
	\label{fig:kick-model} 
\end{figure}

 We apply a CSR point-kick model to analyze the steady-state CSR effect.% This model is oriented toward analyzing the CSR effect . 
As introduced in Ref. \cite{Dohlus:1998}, 
the CSR effect induced in a dipole can be formulated equivalently with a point-kick at the center of the dipole. The idea is illustrated in Fig. \ref{fig:kick-model}. Such point-kick leads to a  horizontal coordinate deviation in $(x,x^\prime)$  plane and an energy deviation $\delta$, which   have  the form of   \cite{Jiao:2014gja}
\begin{equation} 
	X_k= 	\left[   \begin{array}{c  }
		x_k   \\
		x_k^\prime
	\end{array}   \right]=
	\left[   \begin{array}{c  }
		\rho^{4/3} k[\theta \cos(\theta /2)-2 \sin(\theta /2) ]  \\
		\sin(\theta /2) (2 \delta+\rho^{1/3} \theta k )
	\end{array}   \right],
\end{equation}

\begin{equation} 
\delta =  \delta_0+ \delta_{csr} = \delta_0+  k  \rho^{1/3} \theta. 
\end{equation}
  Here the $\delta_0$ and $\delta_{csr}$ indicate the particle’s initial energy deviation and   CSR-induced energy deviation in the upstream dipoles, respectively. 
 The  $\delta_{csr}$ increases by $k\rho^{1/3} \theta$ right after experiencing the kick. 
The parameter $k$ is relevant to the   Gaussian bunch  as \cite{Jiao:2014gja}
\begin{equation}\label{eq:kk}
	k=0.2459\frac{N_b r_e}{\gamma \sigma_z^{4/3}},
\end{equation}
where $N_b$ is the electron population, $r_e$ is the classical electron radius,   $\gamma$ is the   relativistic Lorentz factor. The rms bunch length $\sigma_z$, as a function of  $ s $, is the varying bunch length.
%after passing through the  dipole center.     
 For the following chicane calculation using the point-kick model, 
we assume that $\sigma_z$ and the corresponding $k$ are constant within a dipole and that each dipole has a different $\sigma_z$ and $k$. 
 Specifically, the bunch lengths in the 1st and 4th dipoles  are equal to  the  bunch lengths at the chicane entrance and exit, respectively;  the bunch lengths in the 2nd and 3rd dipoles are equal to the bunch lengths  at the centers of the dipoles. 

The emittance growth due to CSR can be suppressed  by minimizing the  CSR-induced coordinate shifts relative to the reference trajectory at the chicane exit. 
  This point is easily obtained  from  the expression of the transverse emittance in the presence   at the exit  of the beamline:  \cite{Emma:1997hj} 
 \begin{equation}\label{eq:em}
 	\varepsilon_x^2= \varepsilon_{x0}^2+\varepsilon_{x0}\left(\beta_x \left\langle\Delta x_f^{\prime 2}\right\rangle +2 \alpha_x\left\langle\Delta x_f \Delta x_f^\prime \right\rangle +\gamma_x \left\langle  \Delta x_f^2 \right\rangle   \right)+ \Delta \varepsilon_{x }^2.
 \end{equation}
 Here $\Delta \varepsilon_{x }^2= \left\langle\Delta x_f^2\right\rangle \left\langle\Delta  x_f^{\prime 2} \right\rangle-\left\langle\Delta x_f  \Delta x_f^\prime  \right\rangle^2=  0 $. Since the particle coordinate deviations   $\Delta  x_f $ and $\Delta x_f^{\prime}  $ from CSR field are correlated, this term goes to 0   \cite{Venturini:2016dqu}.  Here $\varepsilon_{x0}$ is the  geometric emittance at the entrance $s_0$, and  $ \alpha_x,\beta_x,\gamma_x$  are the Twiss functions at the beamline exit.

Now we perform the 2D point-kick analysis of the four-bend chicane.  
To simplify the analysis, 
we assume a zero initial energy deviation of $\delta_0=0$ and  a zero initial particle coordinates of $X_0=(x_0, x_0^\prime)^\dagger =(0,0)^\dagger $ at the chicane entrance.
After passing through the section from the center of 1st dipole to the center of 2nd dipole,  the particle experiences the second kick,  
\begin{equation}\label{eq:xz2}
	X_{s2}= M_{c1c, 2\times 2}X_{k1}+ X_{k2}.
\end{equation}
%\begin{figure*} 
%\begin{equation} \label{eq:xz2}
%	X_{z2}= M_{12, 2\times 2} X_{k1}+	X_{k2}=	 
%	\left[   \begin{array}{c  }
%		 k_1\rho_1^{4/3} (-2S_1 +C_1 \theta_1) +k_2\rho_2^{4/3}  (-2S_2 +C_2 \theta_2)+S_1  (\frac{1}{2}k_1 \rho_1^{1/3}  \theta_1+ \delta_0)   ( 2L_{d1}+ S_1 \rho_1+  S_2 \rho_2 )     \\
%2 S_1 \delta_0+2 S_2 \delta_1+k_1	S_1  \theta_1 	\rho_1^{1/3}  +k_2	S_2  \theta_2 	\rho_2^{1/3}  
%	\end{array}   \right]
%\end{equation}
%\end{figure*}
From the $2\times 2 $ transfer matrix of the horizontal betatron motion  in Eqs. \eqref{eq:RB} and \eqref{eq:Ld}, the transfer matrix $M_{c1c, 2\times 2}$  between the center of the first two dipoles is given by
\begin{equation}\label{eq:m1222}
	M_{c1c, 2\times 2} =( R_{HB2} R_{d1}  R_{HB1})_{2\times 2}=  \left[ \begin{array}{c  c}
		1& L_{d1} + ( \theta_1 \rho_1  +   \theta_2 \rho_2)/2   \\
		0 & 1
	\end{array} \right], 
\end{equation}
where  $R_{HB1}$ and $R_{HB2}$ are the transfer matrix of  the 1st and 2nd  half-dipoles, respectively. 
Note that  Eq. \eqref{eq:m1222} is a  universal expression for  (half dipole)+drift+(half dipole) structure,  where one just needs to change the subscripts to obtain the matrix of other similar structure. 

Similarly, the particle coordinate deviations after the 4th kick are    
\begin{equation}\label{eq:xz4} \begin{aligned} 
		X_{s4}&= M_{c3c, 2\times 2}X_{s3}+ X_{k4} \\
&=M_{c3c, 2\times 2}( M_{c2c, 2\times 2}X_{s2}+ X_{k3} ) + X_{k4}\\
		&= M_{c3c, 2\times 2}M_{c2c, 2\times 2} X_{s2}+ M_{c3c, 2\times 2} X_{k3}+ X_{k4}.	\end{aligned}
\end{equation} 
The description of $M_{c2c, 2\times 2}$ and $M_{c3c, 2\times 2}$ can be obtained by substituting the  $\rho_1,\rho_2,  \theta_1, \theta_2, L_{d1}$ in Eq. \eqref{eq:m1222} with  the     counterparts      $\rho_2,\rho_3,  \theta_2, \theta_3, L_{d2}$ and  $\rho_3, \rho_4,  \theta_3, \theta_4, L_{d3}$, respectively. 
 Note that the net energy deviation  $\delta_j$   increases by $ k_j \rho_j^{1/3} \theta_j$ after passing through the $j$th dipole, which can be written as  
 \begin{equation}\label{eq:deltai}
\begin{aligned}
	&\delta_1=   k_1 \rho_1^{1/3} \theta_1, ~~~ \delta_2= \sum_{i=1}^{2} k_i \rho_i^{1/3} \theta_i, \\
  &\delta_3= \sum_{i=1}^3 k_i \rho_i^{1/3} \theta_i  ,~~~ \delta_4= \sum_{i=1 }^4 k_i \rho_i^{1/3} \theta_i  ,
\end{aligned}
\end{equation} respectively. 
Finally, the particle coordinate deviations $X_f$ at the chicane exit are 
\begin{equation}\label{eq:1111}  
		X_{f}= ( R_{HB4} )_{2\times 2}   X_{s4}=\left[
\begin{array}{c  } 
 X_{s4}(1,1)+ \frac{\theta_4 \rho_4}{2}  X_{s4}(2,1) \\
	X_{s4}(2,1)
	\end{array} \right],
\end{equation}
 %The   can be expressed as $  \Delta x_f   =    X_{f}(1,1) $, $\Delta x_f^{\prime  }=  X_{f}(2,1)$,  respectively. 
where $ X_{s4}(1,1),  X_{s4}(2,1)$ are the elements of $X_{s4}$  in Eq. \eqref{eq:xz4}. Therefore, the CSR-induced emittance growth at the chicane exit can be theoretically cancelled when the  particle coordinate deviations satisfy $X_{f}=(0,0)^\dagger$, which can also be written as  $X_{s4}=(0,0)^\dagger$.

Although the particle coordinate deviations in the chicane exit are obtained, it is of considerable complexity of the obtained $X_{s4}$. Therefore, a Taylor expansion  is used  to simplify  the sine and cosine terms in $X_{s4}$ with respect to the dipole bending angle $\theta_i ~(i=1,2,3,4)$, 
employing a small bending-angle approximation. 
Besides, we  neglect  the lengths of the dipoles   as $L_{B }\ll L_{d }$.
	Whereby  two analytical CSR cancelation conditions can be obtained in a greatly simplified form  with the aid of the achromatic condition (Eq. \eqref{eq:achromatic-condition}) as 
\begin{equation}  \label{eq:condition1}
	\frac{ q_2  }{q_3 q_4 \ell_3 } =\frac{ \delta_4-\delta_2 }{\delta_2}, ~ 
\ell_2= -\frac{1}{q_3} \frac{    q_3(\delta_3-\delta_1)+(q_2+q_3) \delta_2 }{   \delta_2  +( 1+q_2)(\delta_3-\delta_1)} ,
\end{equation}
where $q_2=\theta_2/\theta_1, q_3= \theta_3/\theta_1, q_4= \theta_4/\theta_1,\ell_2= L_{d2}/L_{d1},\ell_3= L_{d3}/L_{d1} $. 
%Note that Eq. \eqref{eq:condition1} is still complicated because the $k_i ~(i=1,2,3,4)$ in  Eq. \eqref{eq:deltai} is quite   tedious (or use cumbersome?).  

Although the CSR cancelation conditions in Eq. \eqref{eq:condition1} look  complicated,  %(the $k_i ~(i=1,2,3,4)$ in  $\delta_j ~(j=1,2,3,4)$ \textcolor{red}{is quite tedious, or cumbersome?}), 
 they still reveal some important information.  
 First, Eq.  \eqref{eq:condition1} enable us to prove that a symmetric C-chicane  seems connot   fully cancel the net CSR point-kick. Specifically, combined with these conditions for a symmetric C-chicane:  $  q_2=q_3=-1, q_4 = 1,   \rho_1=-\rho_2=-\rho_3=\rho_4$, and $\ell_3=1$,  the first member of Eq. \eqref{eq:condition1}  can be reduced to   $ 	k_1 +k_2 = k_3  +k_4 $. 
This condition cannot be satisfied, because $k_i$ in Eq. \eqref{eq:kk}   decreases as $i$ increases during the compression process.  %What is more, $k_i$ is postive, so   both conditions in Eq. \eqref{eq:a1}  can not be satisfied.
Second,  Eq.  \eqref{eq:condition1} also indicates that less than four dipoles is impossible  to satisfy the CSR cancelation conditions.  This is because,  a chicane with three different dipoles,  can be regarded as a four-bend chicane with one dipole's bending angle set to zero.  
Note that the first member of Eq. \eqref{eq:condition1}  cannot be satisfied  regardless of arbitrary $q_i ~(i=2,3,4)$ set to zero.     Self-evidently, 
 one dipole can cause the CSR effect, and a compressor consisting of two different dipoles is not  achromatic.  

In total,  it can  be observed that there are  \textit{seven}  degrees of freedom  in a chicane, including $q_2, q_3, \rho_2/\rho_1, \rho_3/\rho_1, \rho_4/\rho_1, \ell_2  $ and the compression factor $C$ hidden in $k_i$.
%, are five more than the number of constraints available (only two conditions as expressed in Eq.  \eqref{eq:condition1}). 
The  not mentioned  quantities  $ q_4, \ell_3$ can be obtained from Eq. \eqref{eq:achromatic-condition}. Additionally,  we need not care much about the values of the $\theta_1,\rho_1$ and $L_{d1}$ here, because  the chicane design goals (including the total length $L_{tot}$, the 1st dipole length $L_{B1}$ and momentum compaction $	R_{56}^{s_0\to s_f}$) impose limits on the three variables. 
And the specific chicane layout can be realized by adjusting the
ratios of other bending angles, other bending angles, and other drift lengths to  $\theta_1,\rho_1$ and $L_{d1}$, respectively. 
To illustrate the CSR cancelation conditions quantitatively,  it is necessary to reduce the \textit{seven} degrees of freedom to \textit{three}. Then one can derive the chicane parameters as a function of compression factor $C$    under the \textit{two} CSR cancelation conditions  in  Eq.  \eqref{eq:condition1}. Here we choose conpression factor $C$ as the independent variable in order to ensure the  general applicability of our study.    % we find that the chicane contains excessive number of  degrees of freedom.  To provide constraints on the chicane design,  several conditions are added,
From the  above analysis, we see that we need include \textit{four} additional conditions for chicanes. Indeed,  adding \textit{four} conditions is not difficult because \textit{three}    conditions can be achieved just by ensuring that the four dipoles have either equal bending radii or equal dipole lengths. 
These two options are equivalent for the purpose of adding three conditions. 
 %From this point, the primary factor causing the difference  between Secs. \ref{C-}  and \ref{S-} is the sole remaining condition. 

\section{Application to four-bend C-Chicane}\label{C-}

\begin{figure}[h]
	\centering
	\includegraphics[scale=0.16]{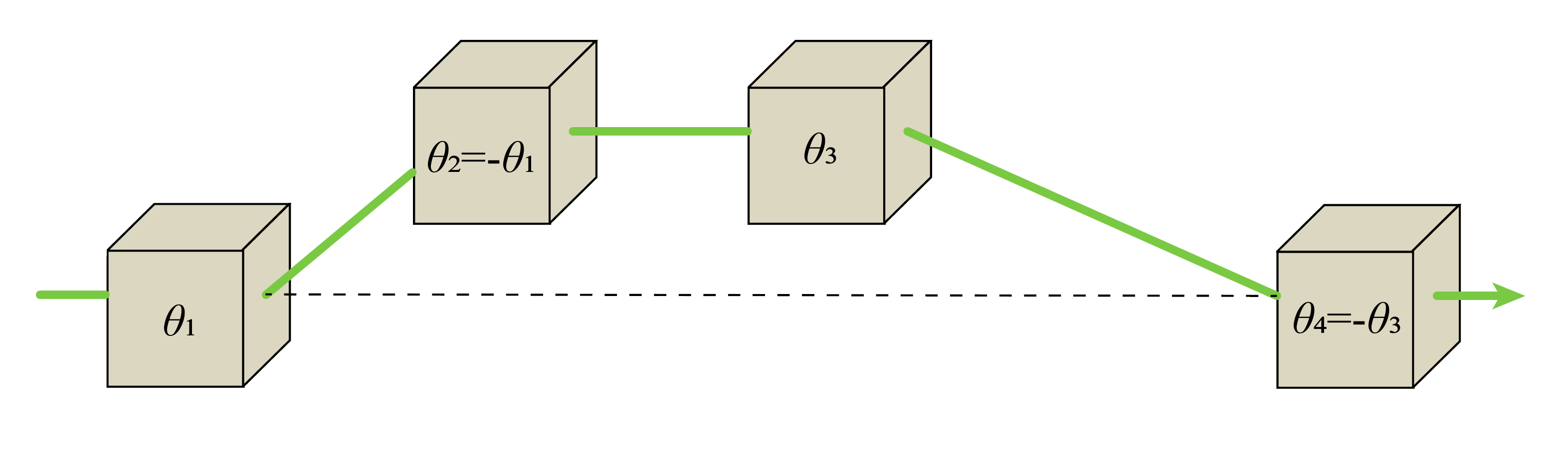}
	\caption{  General geometry of an asymmetric 
		C-chicane. The space between the 2nd and 3rd dipoles is parallel to the entrance or exit of the beam, which is a ``negative drift'' constructed by several quadrupoles.  } 
	\label{fig:lattice} 
\end{figure}

Currently, the symmetric C-chicane is the most commonly used  bunch compressor in operation or under construction.  
Thus we hope to discuss the implementation of a chicane based on a symmetric C-chicane  as much as possible. At this point, we make the bending angles of the 1st dipole and 2nd dipole identical except for the bending direction (denoted as $\theta_1=-\theta_2$ or $q_2=-1$).  
The schematic layout is shown in Fig. \ref{fig:lattice}. This condition $q_2=-1$ preserves  the parallelism of the beam orbits between the 2nd and 3rd dipoles,  indicating a similarity with the symmetric C-chicane.  The condition $q_2=-1$  also results in  $\theta_3$ being equal to $-\theta_4$.  This   greatly simplifies  the structure of the chicane,  reducing the four different dipoles to only two types.    
 %The condition ($q_2=-1$) also results in  $\theta_3$ being equal to $-\theta_4$, the structure of the chicane,   as the four different dipoles are now reduced to two types of dipole.    and their  bending angles ratio $q_3=\theta_3/\theta_1$ provides a clear sign for  better understanding the CSR compensation later. 
Importantly,  the condition $q_2=-1$ also greatly simplifies the achromatic condition and CSR cancelation conditions. First, the achromatic condition in  Eq. \eqref{eq:achromatic-condition} can rewrite as
\begin{equation}  \label{eq:q2q3l3} 
  q_4=-q_3, ~~~ \ell_3=- 1/q_3.
\end{equation}
Substituting  the  above Eq. \eqref{eq:q2q3l3}  in Eq. \eqref{eq:condition1}, 
the   CSR cancelation conditions can be reduced to %simplified to 
\begin{equation}  \label{eq:condition3}
	-\frac{ 1 }{q_3   } =\frac{ \delta_4-\delta_2 }{\delta_2},~~
\ell_2= -  \frac{  \delta_3-\delta_1 }{    \delta_2   }  +  \frac{      1 }{   q_3    }-1. 
\end{equation} 
Note that the second member of Eq. \eqref{eq:condition3}  turns out that  $\ell_2= L_{d2}/L_{d1}$ is negative due to the negative   $q_3$ and the positive $ \delta_3-\delta_1,\delta_2 $. 
Despite the presence of ``negative drift'', we note that the dispersion and the beam trajectory remain  unaffected, since Eqs. \eqref{eq:achromatic-condition-all} and   \eqref{eq:achromatic-condition} show the independence  of $L_{d2}$ under the condition of  $q_2=-1$.
Indeed, this ``negative drift'' is not uncommon in accelerator physics and can be realized by using a focus section with several quadrupoles and drifts \cite{cite:Chao}.    
 For clarity, 
 in this section we use $L_{d2}^{\text{real}} (\ell_2^{\text{real}})$ to describe the real length of the physical space, and $L_{d2}^{\text{eff.}} (\ell_2^{\text{eff.}})$ for a focus section whose transfer matrix is  the same as that of a  ``negative drift''.

Together with the condition  $q_2=-1$,  two typical cases of fixed bending radii and fixed dipole lengths are discussed to provide \textit{four}  conditions,  which can be expressed  as
\begin{equation}  \label{eq:case-1-2}
	\begin{aligned}
\text{Case 1: }& q_2 =-1,~ \rho_2/\rho_1=-1,~ \rho_3/\rho_1=-1, ~  \rho_4/\rho_1=1. \\
\text{Case 2: }&    q_2 =-1, ~ \rho_2/\rho_1=-1 ,~ \rho_3/\rho_1=1/q_3, ~  \rho_4/\rho_1=-1/q_3.
\end{aligned}
\end{equation} 
These two cases correspond to two scenarios: 
Case 1 is given priority during the chicane design phase,  where the bending radii are fixed by setting the equal magnetic field $B_0$; Case 2 is typically used for pre-manufactured dipoles, allowing for adjustment of the dipole's bending angles by changing   $B_0$. 

 %In order to put our theoretical calculation into practice, two typical cases are discussed to  elaborate  the	CSR-cancelation conditions in Eq. \eqref{eq:condition3}.  
 
\subsection{ CSR-immune asymmetric C-chicane }\label{sub:C}

%In the conventional design of a  symmetric  $C$-chicanes, the drift between the 2nd and 3rd dipoles is parallel to the beam entrance, which offers room for hosting beam diagnostics, beam position monitors, and scrapers or masks for beam shaping \cite{M:2015Schreck,Borland:2000ah}. 

%Actually, the negative drifts, consisting of several quadrupoles and drifts, are always used in dealing with path length dynamics \cite{cite:Chao}. 

%It is not hard to find that both cases imply that dipoles 1 and 2 are the same, and dipoles 3 and 4 have the same strength.
% Here we consider   producing as few types of magnet  dipoles as possible   to reduce magnet design and production costs. 

\begin{figure}[ht] 
	\includegraphics[scale=0.26]{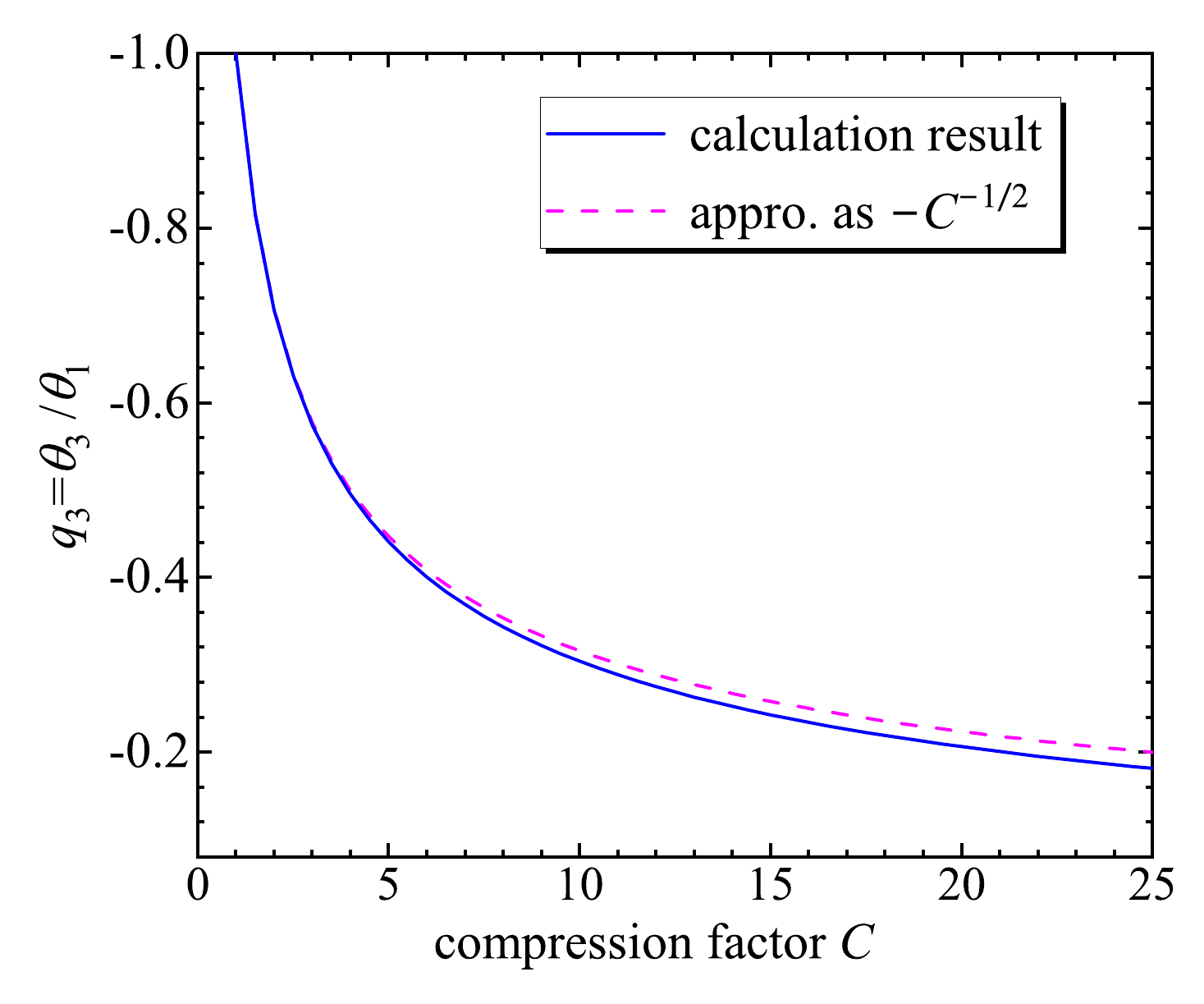}
	\includegraphics[scale=0.33]{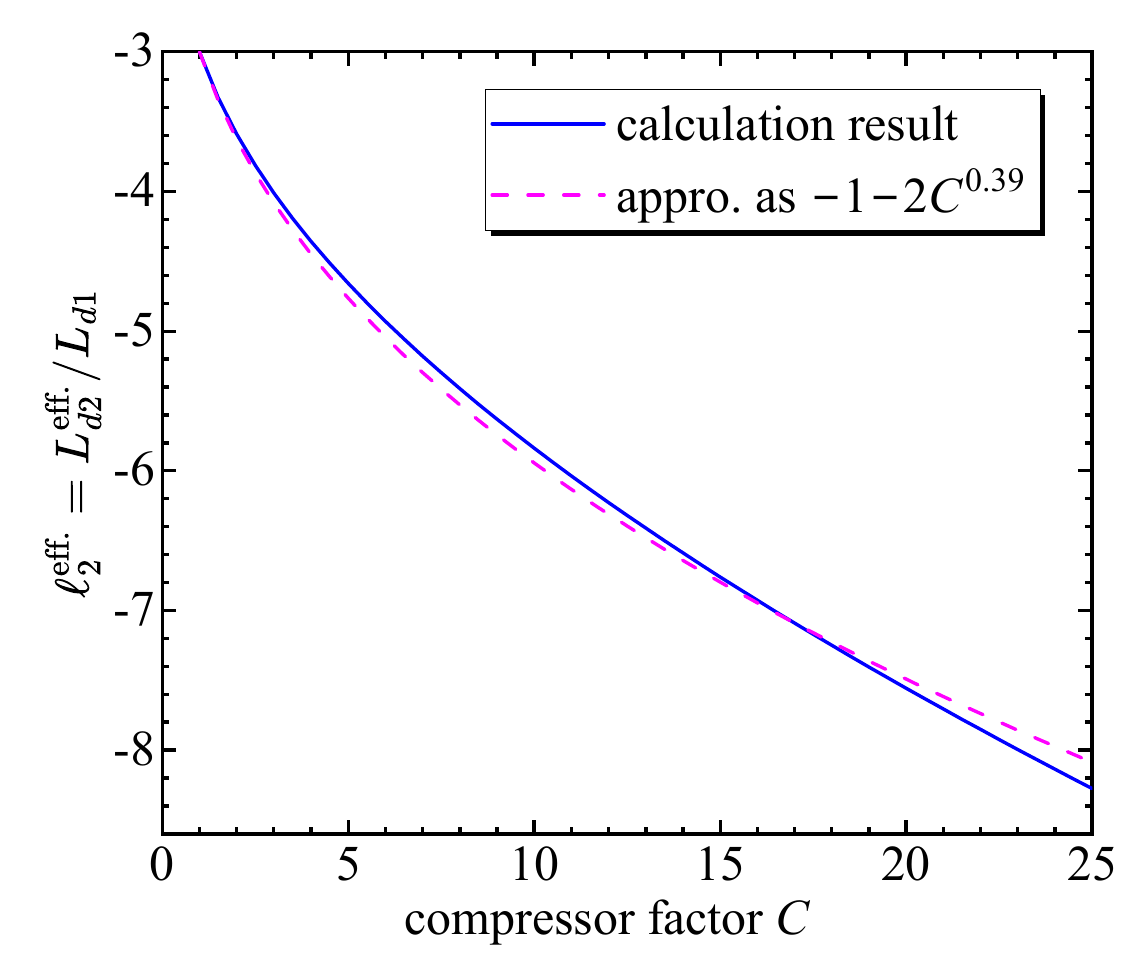}
	\caption{  The required  bending angles ratio  $q_3=\theta_3/\theta_1$  (upper) and  ratio   $\ell_2^{\text{eff.}}=L_{d2}^{\text{eff.}}/L_{d1}$  (bottom) vary with the compression factor $C$ from CSR cancelation conditions  Eqs. \eqref{eq:q2} and \eqref{eq:simpm12}  (blue solid   curves) for Case 1. 
The pink dashed  curves show simpler expression   $q_3 \approx -C^{-1/2} $ and  $\ell_2^{\text{eff.}}  \approx -1-2C^{0.39} $, respectively.  }
		%top (bottom)
	\label{fig:same-rho}  
\end{figure}

\textit{For Case 1}, bending radii are  fixed value as  $    |\rho_1| =|\rho_2|=|\rho_3|=|\rho_4|$.  
By substituting the $k_i ~(i=1,2,3,4)$ (as presented in Appendix \ref{appendixki}) in Eq. \eqref{eq:condition3}, 
%Indeed, the main challenge in calculating the CSR cancelation conditions is the complexity of $k_i$, especially the $k_2$ and $k_3$ (see Appendix \ref{appendixki}).   
   the CSR cancelation conditions for a CSR-immune chicane can be written as
\begin{equation} \label{eq:q2} 
	C^{-4/3}+\left[ \frac{   2 (1-q_3)   }{    1+C -2 C q_3    }  \right]^{4/3}     =q_3^2 \left\{  1+ \left[ \frac{ 2 (1-q_3) }{  2-q_3-C q_3   } \right]^{4/3} \right\}, 
\end{equation}
and
\begin{equation}\label{eq:simpm12} 
\begin{aligned}
\ell_2^{\text{eff.}}   =&   -1+ \frac{1}{ q_3}    -  \frac{    \left[ 2C  (1-q_3)  \right]^{4 / 3}  }{   (  C  - 2  Cq_3+ 1        )^{4/3}+\left[  2C  (1-q_3)    \right]^{4 / 3}    }    \\
&  +\frac{ 2^{4 / 3}        }{ q_3\left\{  \left[ \frac{2 -(1+C)q_3 }{ 1-q_3 }\right]^{4/3} +  2^{4 / 3}  \right\}     } .
\end{aligned}
\end{equation}  
Since Eqs. \eqref{eq:q2} and \eqref{eq:simpm12} are too complicated for  analytical solutions,  one can obtain the numerical results of $q_3$ vs.  $C$ and  $\ell_2^{\text{eff.}}$ vs. $C$ from Eqs. \eqref{eq:q2} and \eqref{eq:simpm12}.  
%One can solve Eq. \eqref{eq:q2} to obtain a relationship between the bending angles ratio  $q_3$ and the compression factor $C$. 
The corresponding quantities  $q_4$ vs.  $C$ and $\ell_3$ vs.  $C$ can be found from the solved $q_3$ vs.  $C$,  since  $q_4=-q_3,   \ell_3=- 1/q_3.$ 
  Figure \ref{fig:same-rho}  shows that the solved  $q_3 $ ($\ell_2^{\text{eff.}}$) is equal to $-1$ ($-3$) for no compression case ($C= 1$),  and   decreases as the conpression factor $C$ increases. Indeed, approximate but explict expressions of $q_3$ and $\ell_2^{\text{eff.}}$  can also be obtained by fitting the  numerical results  for factor $C<25$ as 
\begin{equation} \label{eq:q3l2} 
	q_3\approx -C^{-1/2}, ~~~\ell_2^{\text{eff.}}\approx -1-2C^{0.39}.
\end{equation} 
The comparison between the approximate and theoretical results is shown in Fig. \ref{fig:same-rho} by the dashed and  solid   curves. 
 % The excellent contrast results shows that the  tedious expression in  Eq. \eqref{eq:q2} can be replaced by $q_3 \approx -C^{-1/2} $, which must satisfy the appetite of the accelerator practitioners.
Importantly, 
Eq. \eqref{eq:q2} (or Fig. \ref{fig:same-rho}, upper plot) indicates that shortening the dipole lengths of the last two dipoles is necessary to achieve a CSR-immune chicane. 
This point can be understood qualitatively as the bunch length is shorter at the last two dipoles, where CSR is stronger.

%, and the comparisons  are shown in the solid  blue and dashed pink curve in Fig. \eqref{fig:same-rho}.   

\textit{For Case 2},  the dipole lengths satisfy $    L_{B1} = L_{B2}= L_{B3}= L_{B4} $. %, and the bending radii ratio   $q_3=- \theta_3/ \theta_1 =- \rho_1/ \rho_3 $ are adjusted to achieve the CSR-cancelation condition. 
 Similarly to Case 1,   we obtain the empirical equations of  $q_3$ and $\ell_2^{\text{eff.}}$  for  the factor $C<25$ as 
\begin{equation} \label{eq:q4} 
	 q_3 \approx -C^{-3/5}  , ~~~\ell_2^{\text{eff.}}\approx -1-2C^{0.47}, 
\end{equation} 
for the case of fixed dipole lengths. 
%Note that Eqs. \eqref{eq:q2}  and \eqref{eq:q4}  all reveal one conclusion that a CSR-cancelation chicane with four same dipoles has no compression,  as shown in Figs. \eqref{fig:same-rho} and   \eqref{fig:same-L}.  
 %The values of $q_3$ and $\ell_2$, calculated from   CSR-cancelation condition, are compared with the simple expression in Eq. \eqref{eq:q4},  which illustrates an agreement. 
From Eq. \eqref{eq:q4}, it can be seen that the key to achieving a CSR-immune chicane, as alreadly described in  Ref. \cite{Stulle:2007se}, is to weaken     the last two dipoles compared with a traiditional symmetric C-chicane. 
However, the parameter scans and particle tracking approach in Ref. \cite{Stulle:2007se}  works only in a case-by-case sense. Here we  present  a general demonstration rather than focusing on a practical design.

\begin{figure}[ht] 
\centering
\includegraphics[scale=0.24]{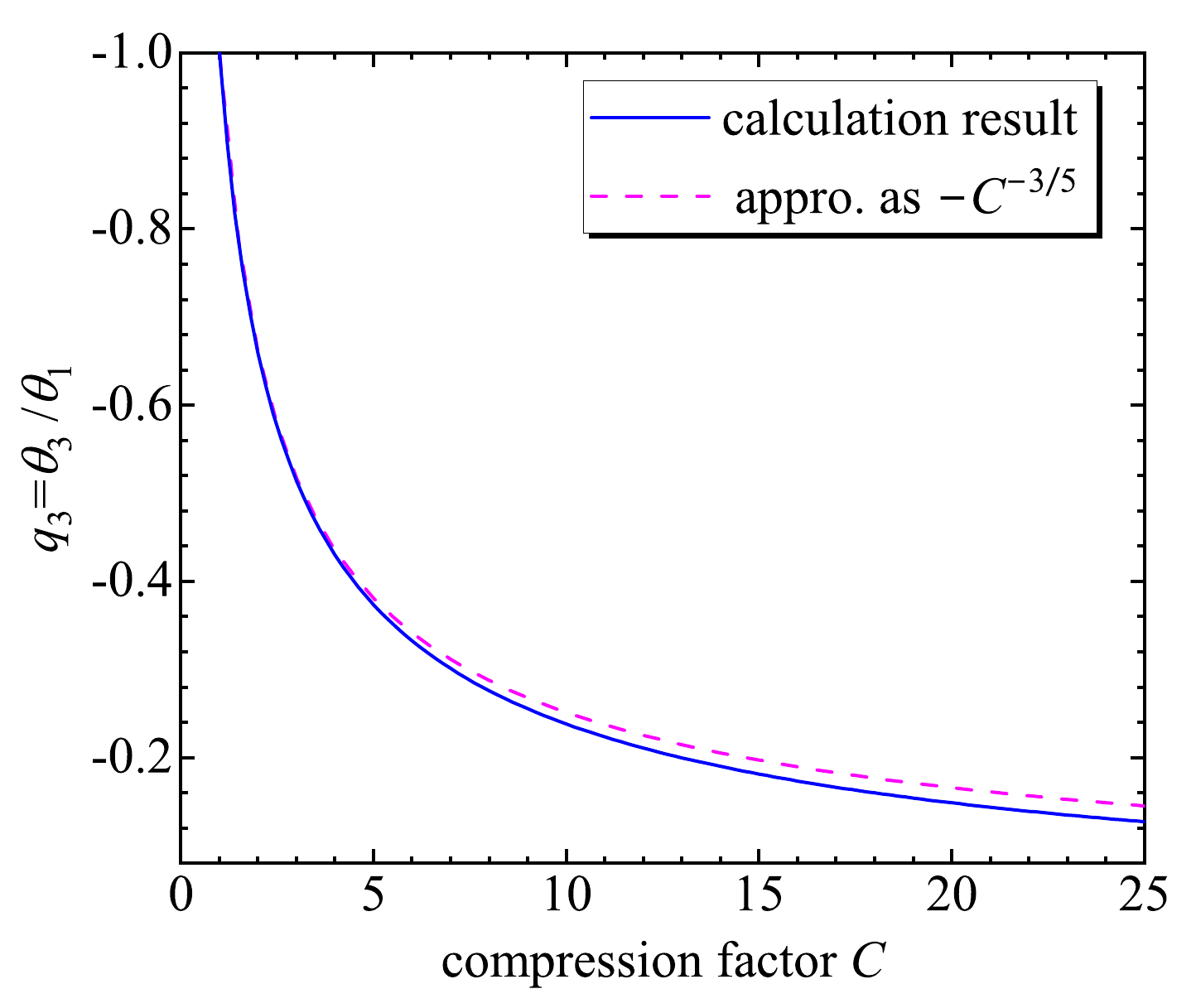}
\includegraphics[scale=0.3]{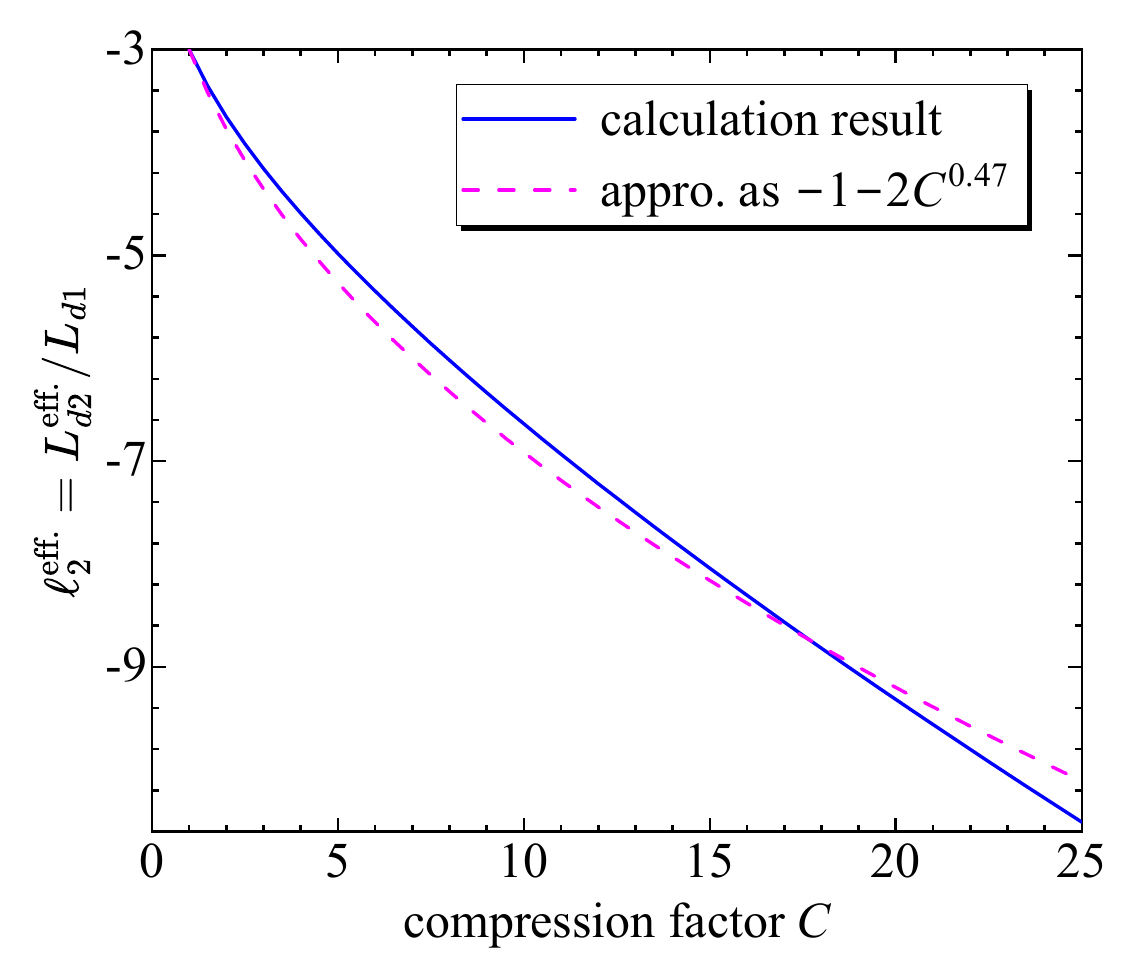} 
 \caption{  The required  bending angles ratio  $q_3=\theta_3/\theta_1$   (upper) and  ratio   $\ell_2^{\text{eff.}}=L_{d2}^{\text{eff.}}/L_{d1}$  (bottom) vary with the compression factor $C$ from CSR cancelation conditions  for Case 2 (blue solid   curves). The pink dashed  curves show   simpler expression    $q_3 \approx -C^{-3/5} $ and $\ell_2^{\text{eff.}} \approx -1-2C^{0.47} $, respectively.  }
	\label{fig:same-L} 
\end{figure} 

% In principle, the proposed design has the same overall structure as the symmetric  $C$-chicane except  (i) the deflection magnitude (dipole length) of the last two dipoles is weaker (shorter) than the first two  with a ratio of $q_3$.  (ii) the distance between the 3rd and 4th dipoles is increased until the ratio $\ell_3=-1/ q_3$ is satisfied.   (iii)   a focus section  matched the ``negative drift" $L_{d2}^{\text{eff.}}$   is provided between the 2nd and 3rd dipoles, consisting of  several quadrupoles. Besides,  the two mentioned values  $q_3$ and $L_{d2}^{\text{eff.}}=L_{d1} \ell_2^{\text{eff.}}$, which only depends on the compression factor $C$, can be empirically expressed through Eqs. \eqref{eq:q3l2}  and \eqref{eq:q4}.  

 Finally, the chicane design goal,  the   $	R_{56}^{s_0\to s_f}$ and   $L_{tot}$, can be obtained from the value of $L_{d1}$, $\theta_1$, and the provided  $q_3$ (see    Figs. \ref{fig:same-rho} and \ref{fig:same-L}  (upper)) as 
 \begin{equation}\label{eq:goal}
	R_{56}^{s_0\to s_f}= -  L_{d1} \theta_1^2  (1 -q_3  ),~~ L_{tot}  =L_{d2}^{\text{real}}+L_{d1}(1-1/q_3), 
 \end{equation}
which indicates that for a given design goal  and a larger compression factor $C$ (that is, a smaller $|q_3|$),   there will  be a smaller $L_{d1}$ and a larger $\theta_1$, and vice versa.   This conclusion   also applies to the comparison with the traditional C-chicane, which is the case of  $|q_3|$=1. 
Our proposed asymmetric design have  larger $|\theta_1|$, $L_{d3}$ and smaller $|\theta_3|$, $L_{d1}$, compared with   symmetric C-chicane with the same design goal.
%It  is expected that the various chicanes with the same design goal can be compared, including 	 $L_{tot}$ and $R_{56}^{s_0\to s_f}$. 

 \subsection{ Numerical verification of the proposed conditions }\label{mo:C}

 In the above subsection, we have obtained generic CSR cancelation conditions  for an  asymmetric  C-chicane. However, these conditions in Eq. \eqref{eq:condition3} are derived under two assumptions: the variation in bunch length within a single dipole is not considered, and the lengths of the dipoles are neglected. 
Therefore, the main purpose  of this subsection is to investigate the value of $(q_3,\ell_2^{\text{eff.}})$  using  more precise calculations, and then to compare it with the result from the point-kick model. Both numerical integration method and ELEGANT particle tracking simulations are performed. 
%In the presence of these two methods,  we note that it is almost impossible to attain an analytical solution for compression progress. 

To this end, the dipole lengths are further taken into account in the calculation of the parameters of $	R_{56}^{s_0\to s_f}$, $L_{tot}$, and $\ell_{3}$ with aid of the Eq.  \eqref{eq:achromatic-condition-all} (see Appendix \ref{r56-tot}). 
%, the  drift length $L_{d3}$, the momentum compaction $	R_{56}^{s_0\to s_f}$, and total length $L_{tot}$ in Eq. \eqref{eq:goal} can be exported by more complex expressions (see Appendix \ref{r56-tot}). 
 For the convenience of comparison, all chicanes analyzed in this study have been assigned the same values for     $L_{tot}$,  $	R_{56}^{s_0\to s_f}$, $L_{B1}$, and  compression factor $C$. For numerical integration, we need to specify the   concrete value  of  these consistent  quantities  during  calculation. 
We set a  design based on a  symmetric C-chicane with typical parameters  of   equal dipole length of $1 \text{ m}$, the 1st and 3rd drift lengths  of $5.5 \text{ m}$,  the length between the 2nd and 3rd dipoles of 5 m,  bending angles of 3.0$^{\circ} $, and the compression goal of $C=10$. Such chicane parameters enable the derivation of the complete information for a symmetric  C-chicane, as listed in Tables \ref{Table-same-Parameter} and \ref{Table-C-chicane}.  
The   design goals in Table \ref{Table-same-Parameter} are common to all chicanes compared in the following. 
% For the convenience of comparison,   the acquired  values for $L_{tot}$,  $	R_{56}^{s_0\to s_f}$, tegother with the 1st dipole length and compression factor (Table \ref{Table-same-Parameter}), are are consistent among all the chicanes analyzed in this study.

   \begin{table}
\caption{Parameters common to all bunch compressor chicanes compared in this paper. } 
\begin{tabular}{l cccccc }
\hline\hline\noalign{\smallskip}
 	&  Symbol~   &  	~chicane	Parameters ~    & ~Unit  \\ \hline		
 \noalign{\smallskip} 
	~Total length &$L_{tot}$ &  20 & m      \\
  \noalign{\smallskip}
	~First-order momentum &   \multirow{2}*{ $R_{56}^{s_0\to s_f}$  } 	 &  \multirow{2}*{~37.5~}  & \multirow{2}*{mm}    \\
~compaction &&&&\\
 \noalign{\smallskip}
 ~Compression factor  ~ &$C $     &10 &    ---  \\   
 \noalign{\smallskip}
 ~Length of the 1st dipole  ~ &$ L_{B1} $     & 1.0 &    m  \\   
\hline\hline
\end{tabular} 
\label{Table-same-Parameter}
\end{table}

 \begin{table}
 	\caption{Parameters of the symmetric and  asymmetric  C-chicane settings for Case 1  with  units in meters. }    \centering
 	\begin{tabular}{lcccc}
 \hline	\hline \noalign{\smallskip} 
 	& \multirow{2}*{~ Symbol ~}  &  ~symmetric~	     &  asymmetric	  \\   \noalign{\smallskip} 
 	 	&     &   	C-chicane	    &   	C-chicane	   &   \\ \hline	
 \noalign{\smallskip} 	
 	Length of the first two dipoles  &$ L_{B1},L_{B2}$   & 1.0 & 1.0 &     \\  
 \noalign{\smallskip} 
 	Length of the last two dipoles &$ L_{B3},L_{B4}$   & 1.0 &  0.30 &       \\  
 \noalign{\smallskip} 
 	Bending radii of each dipole &$ \rho$  & 18.86   & 10.07 &     \\
 \noalign{\smallskip}  
 	Length of the 1st drift &$ L_{d1}$   & 5.5& 2.17 &    \\  
 \noalign{\smallskip} 
 	Length of the 3rd drift &$ L_{d3}$   & 5.5  & 10.23&     \\  
  \noalign{\smallskip} 
 Effective	value of the 2rd drift & $ L_{d2}^{\text{eff.}}$   &  ---  &-20.65&     \\   
 	\hline	\hline  
 	\end{tabular} \label{Table-C-chicane}
 \end{table}

\begin{figure}[ht] 
 	\centering
 	\includegraphics[scale=0.24]{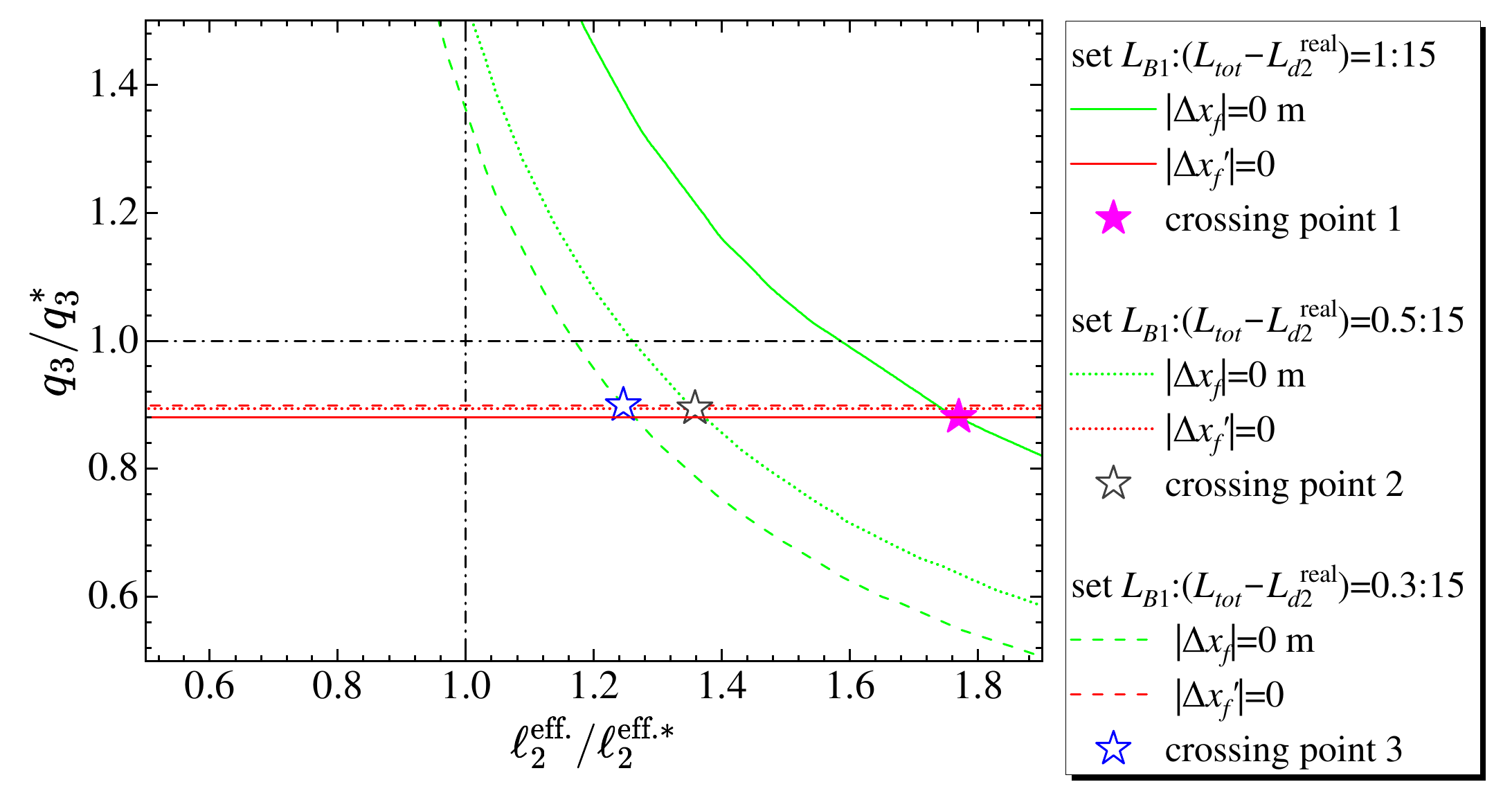}
 	\caption{The $(\ell_2^{\text{eff.}}/\ell_2^{\text{eff.}*}, q_3/q_3^*$) of a  chicane with $|\Delta x_f|=0$ m and $|\Delta x_f^\prime|=0$, as shown in the  green and red  solid curves. The pink  pentacle is the crossing point of the two curves. %, which denotes that both $|\Delta x| $ and $|\Delta x\prime| $ are equal to zero.  
As a comparison, the results for $L_{B1}$:$(L_{tot}-L_{d2}^{\text{real}})= $1.0:15, 0.5:15 and  0.3:15,    are calculated as $(\ell_2^{\text{eff.}}/\ell_2^{\text{eff.}*}, q_3/q_3^*$)=(1.770, 0.881), (1.358, 0.894)   and  (1.246, 0.899), respectively. 
 		%$|\Delta_x|\le 5\times 10^{-8}$ and $|\Delta_x'|\le 1\times 10^{-7}$
 	} 
 	\label{fig:cross} 
 \end{figure}

Now we verify the theoretical result of the point-kick model. 
The CSR-induced coordinate deviations are evaluated  using a numerical integration method  \cite{Emma:1997hj}, which can be considered as a bunch length variation model   within each dipole. 
The verification for our calculation is similar to Ref.  \cite{Zhang:2023cgl}, and the   details  are presented  in   Appendix \ref{appendix1}.
%Taking the case with dipoles of the same bending radii as an example. 
From Eqs. \eqref{eq:q2} and \eqref{eq:simpm12},  we can obtain the theoretical values of  $(q_3^*,  \ell_2^{\text{eff.}*}) =(-0.30,  -5.84)$ for a bunch  compressed by a factor of 10. 
 % To verify these results, some certain chicane parameters need to be specified for the  integration calculation. 
  Here we consider a chicane case with the parameters the same as those listed in   Table \ref{Table-same-Parameter}, and  a physical length of $L_{d2}^{\text{real}}=5 \text{ m}$, which  is designed to accommodate quadrupoles.  
 According to the  integration method, the  calculation results of $(\ell_2^{\text{eff.}}/\ell_2^{\text{eff.}*},q_3/q_3^*) =(1.770, 0.881)$ can be derived from $|\Delta x_f|=0$ m and  $|\Delta x_f^{\prime}|=0$,  as  shown in the pink  pentacle  in Fig.   \ref{fig:cross}.  Here  $\ell_2^{\text{eff.}}$ and $q_3$ are normalized with respect to $\ell_2^{\text{eff.}*} $ and $q_3^*$  just for clear  comparison. This result  $(\ell_2^{\text{eff.}}/\ell_2^{\text{eff.}*},q_3/q_3^*)  $ close to $(1,1) $  indicates a accurate calculation by the point-kick model.  As shown, 
  the  point-kick analysis results and integration results are are slightly different but basically consistent.  

The integration results indicate that  the value of $q_3$ is independent of $\ell_2^{\text{eff.}}$, as shown in the red curves in   Fig.  \ref{fig:cross}. This is not the first time that independence has been achieved, and the previous point-kick analysis also highlighted this situation (see Eq.\eqref{eq:q2}). 
This independence can be attributed to the fact that the expression of $\Delta x_f^{\prime 2}$ in Eqs. \eqref{eq:integrate} does not have $L_{d2}^{\text{eff.}}$ and    depends solely on $q_3$ (see Eqs.  \eqref{eq:r161}-\eqref{eq:r164}). % \textcolor{red}{This is not the first time this has happened, the previous point kick model also shed light on this situation as  $q_3 \approx -C^{-1/2} $ and     $q_3 \approx -C^{-3/5} $ for case 1 and 2, respectively.  }
From this perspective, we find that the CSR-induced coordinate deviation $\Delta x_f^{\prime}=0$ can be achieved simply  by adjusting the deflection magnitude (or dipole length) of the last two dipoles and the first two  with a ratio of $q_3$.  %instead of the ``negative drift'' between the 2nd and 3rd dipoles.  

In the  numerical verification above, we find that the deviations of the obtained $(\ell_{2},q_3)$ with respect to the model calculation, can be attributed to the set ratio between $L_{B1}$ and ($L_{tot}-L_{d2}^{\text{real}}$). As this ratio decreases, $(\ell_2^{ \text{eff.}},q_3)$ become  closer to the theoretical results $(\ell_2^{\text{eff.}*},q_3^*)$, as illustrated in Fig. \ref{fig:cross}. 
 Although  the point-kick analysis requires some fine-tuning,  the differences are minor and can be compensated by parameter adjustments in the actual design process. 
  %Actually,    In fact, we trade this slight loss of accuracy  for a more general CSR cancelation condition. 
 Note that  the integration methods and the following particle tracking, can yield more accurate results, however it is  almost impossible to attain an analytical solution for the compression progress.

\begin{figure}[ht]  
 	\centering
 	\includegraphics[scale=0.22]{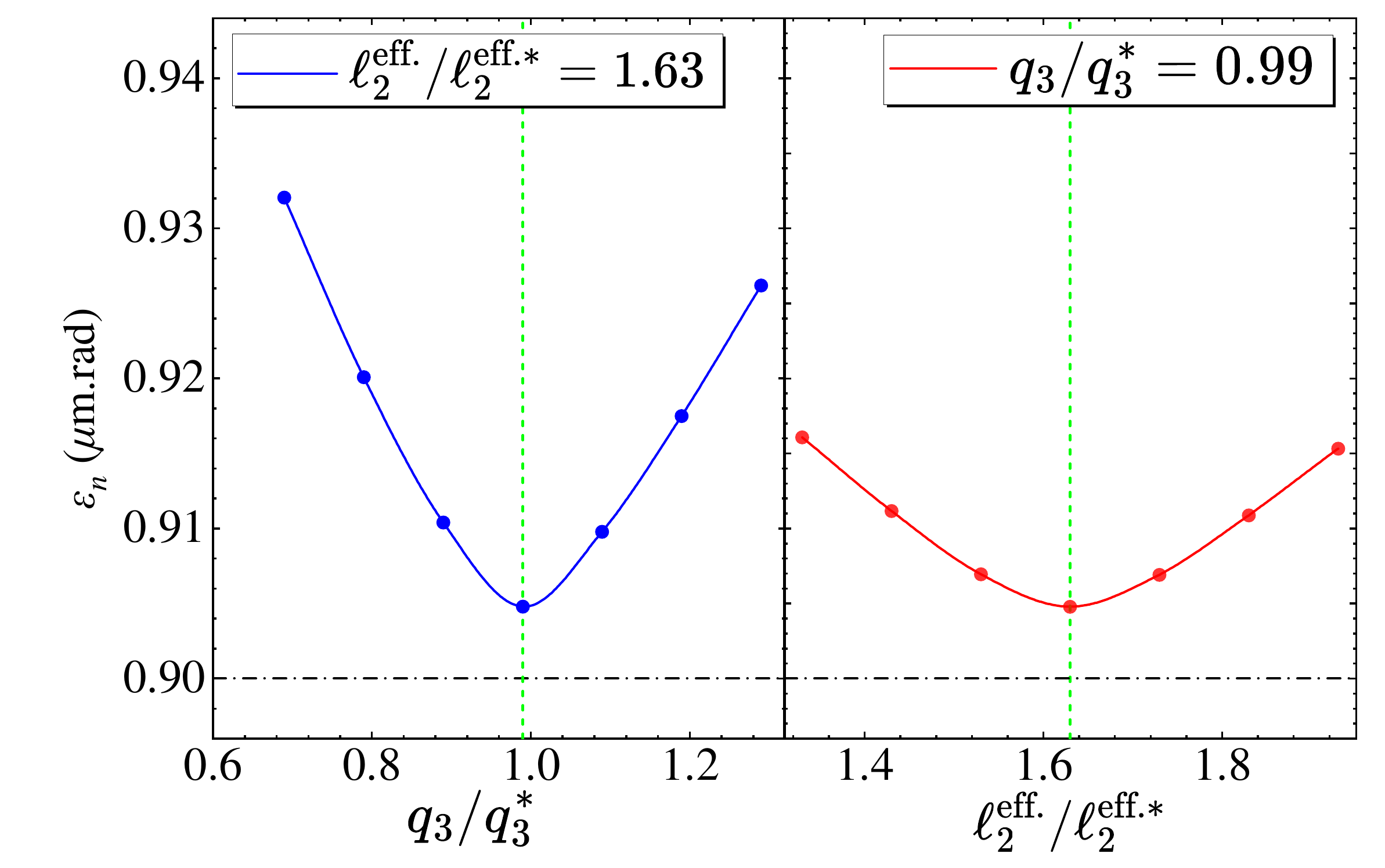} 
 	\caption{  The  ELEGANT simulations result of the  CSR-induced  emittance  $  \varepsilon_n$ near $(q_3^*, \ell_2^{\text{eff.}*})$ for   asymmetric C-chicane.  The simulations results turn  out that the parameters satisfy  $(q_3/q_3^*, \ell_2^{\text{eff.}}/\ell_2^{\text{eff.}*})=( 0.99,1.63)$ with scaned initial Twiss parameters $(\alpha_{x0},\beta_{x0})= (-2 ,10 \text{ m})$, as shown in the green dashed curves.  	}
 	\label{fig:C-chicane-elegant} 
 \end{figure}

On the other hand, to verify the  proposed   CSR cancelation conditions, we simulated the emittance growth caused by steady-state CSR using ELEGANT    \cite{Borland:2000gvh,Borland:2001xua}. The  Gaussian bunch with typical initial parameters is tracked as listed in Table \ref{Table:bunch}. 
The momentum chirp is set as  $h=(1-C)/(C R_{56}^{s_0\to s_f})= 24.02  \text{ m}^{-1}$  from the the values of compressor factor and $	R_{56}^{s_0\to s_f}$ in Table \ref{Table-same-Parameter}. 
%We also set the chicane total length $L_{tot}=20$ m,  a physical length of $L_{d2}^{\text{real}}=5$ m, and the 1st dipole length $L_{B1}=0.5$ m. 
 Here the  initial  Twiss parameters are scanned to  minimum the emittance growth at the chicane exit by optimally matching the beam envelope to the orientation of a nonzero net CSR kick at the  chicane exit \cite{Hajima}.  Fig. \ref{fig:C-chicane-elegant} shows the     simulations result of the  CSR-induced  emittance  $  \varepsilon_n$ near $(q_3^*, \ell_2^{\text{eff.}*})$.
We can see that the minimum  emittance growth $ \Delta  \varepsilon_n/ \varepsilon_n$ is of 0.53\% and the corresponding values are of $(q_3/q_3^*, \ell_2^{\text{eff.}}/\ell_2^{\text{eff.}*})=( 0.99,1.63)$.
These simulation results are close to the results of the point-kick analysis and   basically agree with the  integration results.   
%Here  only the steady-state CSR  is included in the simulation. 
 With aid of the common chicane parameters in Table \ref{Table-same-Parameter} and the scanned  values $(\ell_{2},q_3)$, 
the complete chicane  information can be obtained, as listed in Table \ref{Table-C-chicane}.

 \begin{table}
 	\caption{The initial and final beam parameters for the ELEGANT simulations.} \centering
 	\begin{tabular}{lcccc}
 		\hline	\hline\noalign{\smallskip} 
 		~~~beam	Parameters      & ~~~Symbol~~~   & ~~~Value~~~  & ~~~Unit~~~ &\\ \hline
\noalign{\smallskip} 
 		~~~Bunch charge   &$Q $ & 300   & pC  &   \\
\noalign{\smallskip} 
 		~~~Initial rms	bunch length        & $\sigma_{z0}$           & 100   & $\mu$m  &  \\
\noalign{\smallskip} 
 		~~~Beam energy  &$E_0$    &3  & GeV  &\\
\noalign{\smallskip} 
 		~~~Norm. transv. emittance  & $\varepsilon_{n0}$  & 0.9 & $ \mu $m.rad    &  \\
\noalign{\smallskip} 
 		~~~Rel. rms energy spread & $\sigma_{\delta}$  & 0.01  & \%    &    \\
 		\hline	\hline
 	\end{tabular} \label{Table:bunch}
 \end{table}

%Importantly,  it turns out that the different ratios have an obvious impact on the result of the value of $ \ell_2^{*\text{eff.}}$.

%\textcolor{red}{  To find the reason for the difference between the two methods,} three cases   with different ratio  between $L_{B1}$ and ($L_{tot}-L_{d2}^{\text{real}}$) are studied. %  One main reason is that the verification result by the integration method holds for any $R_{56}^{s_0\to s_f}$ and is only affected by the ratio $L_{B1}$:$(L_{tot}-L_{nd})$.  
   % This is because we assume $L_B \ll L_{d}$ and ignore the length of the dipoles in Eq. \eqref{eq:condition1}.
% Additionally, the small difference between the two methods in $q_3$ arises from our neglect of the variation in bunch length within the dipole.
%Moreover, the small variation in $q_3$ is mainly attributed to bunch length   approximation in point kick model.   

 \section{Application to four-bend S-Chicane   }\label{S-}

 We define the symmetric S-chicane as one in which the  1st and 4th   dipoles have equal dipole length  and the 2nd and 3rd dipoles  are of double that, and all dipoles have the same bending strength,  otherwise it is called asymmetric. 
And the symmetric S-chicane satisfies the position conditions that $L_{tot}=4L_{d1}=2L_{d2}=4L_{d3}$.    
 In this section, we discuss the implementation of an asymmetric  chicane based as much as possible on the  symmetric one, while comparing the  performance of   the symmetric and asymmetric chicanes as a reference for our discussion.
 
\begin{figure}[h]
 	\centering
 	\includegraphics[scale=0.16]{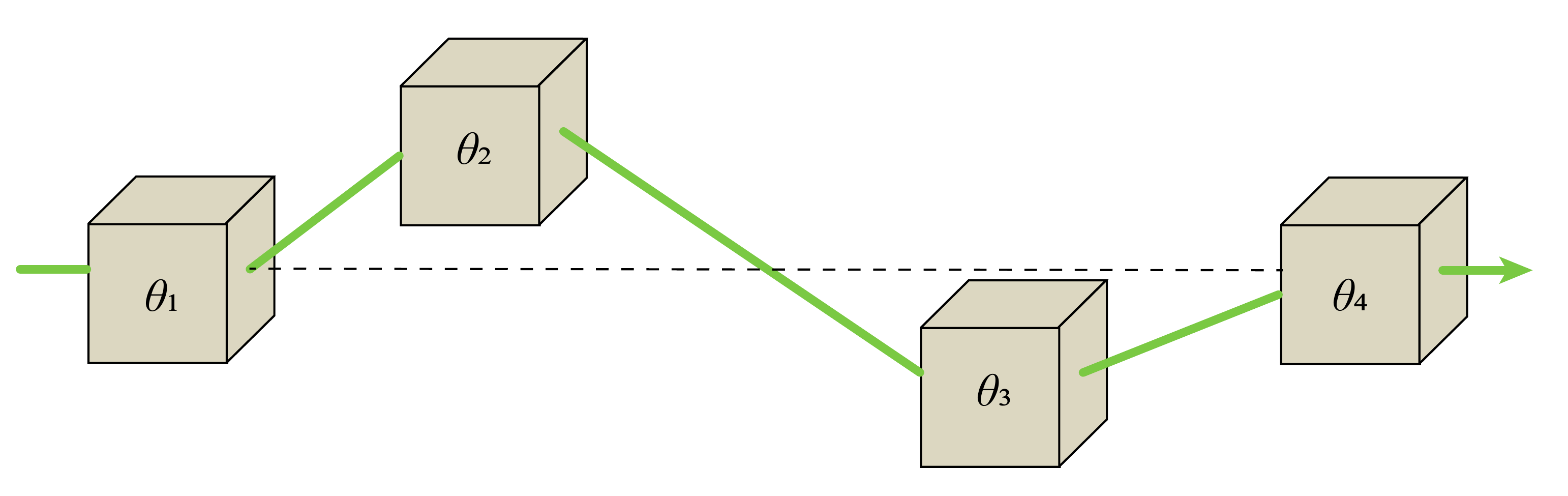}
 	\caption{ General geometry of an asymmetric S-chicane. The drift length between the 1st and 2nd dipoles   is a quarter of the chicane total length.   }
 	\label{fig:lattice2} 
 \end{figure}

 Although S-chicanes are used less frequently than C-chicanes in FELs,   existing studies shows that S-like chicanes exhibit a notable ability to preserve CSR-induced emittance \cite{Antipov:2021eko,Stulle:2007se,Beutner:2007zza,Khan:2022tkg}. 
 The utilization of five or six dipoles is required in such S-like chicanes.   
 In   Sec. \ref{sub:S},   we demonstrate a more ambitious idea of achieving a   CSR-immune chicane using only \textit{four}  dipoles.   
  The  analysis reveals that such chicane has an S-type geometry  as well and  shows that four dipoles are enough to minimize the CSR effects for a chicane (shown in Fig. \ref{fig:lattice2}).   
 
 To further fulfil a CSR-immune chicane,  
 \textit{four}  additional  conditions (as introduced in the end of Sec. \ref{sub:kick}) are  necessary. 
  Firstly, two typical  cases are discussed for chicane designs. Similar to Sec. \ref{C-}, Case 3 and Case 4 denote two chicanes for each dipole with fixed bending radii and fixed dipole lengths, respectively. 
 Nevertheless,  the current provided  \textit{three} conditions are insufficient,  unless  an additional  condition is added. 
 To mimic the structure of a symmetric S-chicane,   
 the ratio $L_{d1}/L_{tot}=1/4$ is set to be similar to a symmetric S-chicane. 
 The ratio $L_{d1}/L_{tot}=1/4$ allows the longitudinal positions of 1st, 2nd and 4th dipoles to be fixed, when the total length is constant. %The position of the 3rd dipole in the longitudinal tegother with these bending angles are adjustable. 
It should be pointed out that,  although not strictly necessary,  we  set this position restrict for practical reasons.  Because  adjusting  dipole positions can be  more difficult than bending angles.  
 Combined with  the achromatic condition in Eq. \eqref{eq:achromatic-condition} and the additional condition $L_{d1}/L_{tot}=1/4$,    the variables $q_4=\theta_4/\theta_1, \ell_2=L_{d2}/L_{d1},  \ell_3=L_{d3}/L_{d1}$   can be expressed in terms of $q_2,q_3$ as 
 \begin{equation}  \label{eq:condition-position}
 	q_4=  -(1+q_2+q_3),~~\ell_2 =\frac{4  }{q_3} +3( \frac{q_2}{q_3}+1),~~\ell_3 =- \frac{4 }{q_3}-\frac{3q_2}{q_3} .
 \end{equation}

 %At this point,  only the third dipole's  shift in the longitudinal direction, together with the strength of four  dipoles,    need  adjust for practical machine, which is a more operational approach than shifting the position of the dipoles.

 Together with the  position condition $L_{d1}/L_{tot}=1/4$,  the \textit{four}  conditions    can be expressed  as
 \begin{equation}  \label{eq:case-1-2}
 	\begin{aligned}
 		\text{Case 3: }& L_{d1}/L_{tot}=1/4,~ \rho_2/\rho_1=-1, ~\rho_3/\rho_1=1,  ~ \rho_4/\rho_1=-1. \\
 		\text{Case 4: }&   L_{d1}/L_{tot}=1/4, ~ \rho_2/\rho_1= 1/q_2, ~  \rho_3/\rho_1=1/q_3, \\
 		&  \rho_4/\rho_1= 1/q_4.
 	\end{aligned}
 \end{equation} 
 The detailed design is  introduced in  Sec. \ref{sub:S-2}. %to provide three constraints,
 The verifications of the CSR cancelation conditions by the integration method and numerical tracking via ELEGANT are given in Sec. \ref{mo:S}.

\subsection{The proof of S-chicane with three positive drifts }\label{sub:S}

The possibility of three positive drifts between dipoles is demonstrated in this subsection.
  The first member of Eq. \eqref{eq:achromatic-condition}  and  the second member of Eq. \eqref{eq:condition1} can be combined to yield   the expression of  $\ell_3$. Thus the expressions of  $\ell_2 $ and $\ell_3 $  can be written as 
\begin{equation} \label{eq:ell-2+3}
	\begin{aligned}
\ell_2=& -\frac{1}{q_3} \frac{ q_3(\delta_3-\delta_1)+(q_2+q_3) \delta_2 }{   \delta_2  +(1+q_2)(\delta_3-\delta_1)}, \\ 
\ell_3 =&\frac{1}{q_3}\frac{ q_2  \delta_2  }{    \delta_2+ ( 1+q_2)(\delta_3-\delta_1)    } 	.    
	\end{aligned}
\end{equation}
Our objective is to ensure that both $\ell_2$ and $\ell_3$ are positive.
Note that $\delta_3-\delta_1$ and $\delta_2$  are both positive. As the signs of  $q_2 $ and $q_3$ are unknown, they can be classified into four groups based on their signs:
  
  \begin{enumerate}[Group 1:]
 \item  set $q_2>0 $ and $q_3>0$.
 This group is not applicable because $\ell_{2}<0$.
 
   \item  set $q_2>0 $ and $q_3<0$.
 This group is not applicable because $\ell_{3}<0$.
 
 \item  set $q_2<0 $ and $q_3<0$. 
     As $\ell_{3}>0$, one can obtain $    \delta_2+ ( 1+q_2)(\delta_3-\delta_1) >0$.
     Thus  $q_3(\delta_3-\delta_1)+(q_2+q_3) \delta_2>0$ as $	\ell_{2}>0$. 
  However,   $q_3(\delta_3-\delta_1)+(q_2+q_3) \delta_2>0$ is impossible as  $q_2<0, q_3<0$.
    
 \item  set $q_2<0 $ and $q_3>0$. The case that $	\ell_{2}>0$ and $	\ell_{3}>0$ is possible under the condition that $q_3(\delta_3-\delta_1)+(q_2+q_3) \delta_2>0$ and $    \delta_2+ ( 1+q_2)(\delta_3-\delta_1) <0$. 
 Thus   the result $ 1+q_2<0$ can be achieved. 
  \end{enumerate} 
Up to now, we have excluded all possibilities based on the taxonomy, except for Group 4.

Lastly, the sign  of $q_4 $ can be determined by reusing the first member of Eq. \eqref{eq:condition1},  which  turns out to be a negative $q_4$. %explicitly. 
 %Another benefit of the negative $q_4$  is that  the value of $|	R_{56}^{s_0\to s }|$  becomes larger and larger after crossing each dipole   (see Appenidx \ref{appendixki}).  
Taken as a whole, a chicane with three positive drifts must satisfy
\begin{equation} \label{eq:s-chicane}
 	q_2<0, ~ q_3>0, ~ q_4<0, ~  1+q_2<0, ~	\frac{\delta_3-\delta_1}{\delta_2}>-\frac{ 1}{ 1+q_2} .
\end{equation}
%In fact, the above classification discussion leads to the same conclusion regardless of the $\theta_1$ set for positive and negative.  
One can find that it can be identified as an   S-chicane (Fig. \ref{fig:lattice2})  from the angle relation.   And the test indicates that the  last condition in Eq. \eqref{eq:s-chicane} is  achievable, e.g., by adjusting the bending radii $\rho_i (i=1,2,3)$ in $\delta_3-\delta_1$ and $\delta_2$. 

%Therefore, using five bend or six bend chicane to canceling the CSR in  \cite{Antipov:2021eko,Stulle:2007se}  seems  unnecessary. 

\subsection{  CSR-immune asymmetric  S-chicane  }\label{sub:S-2}

\begin{figure}[h]
	\centering
	\includegraphics[scale=0.27]{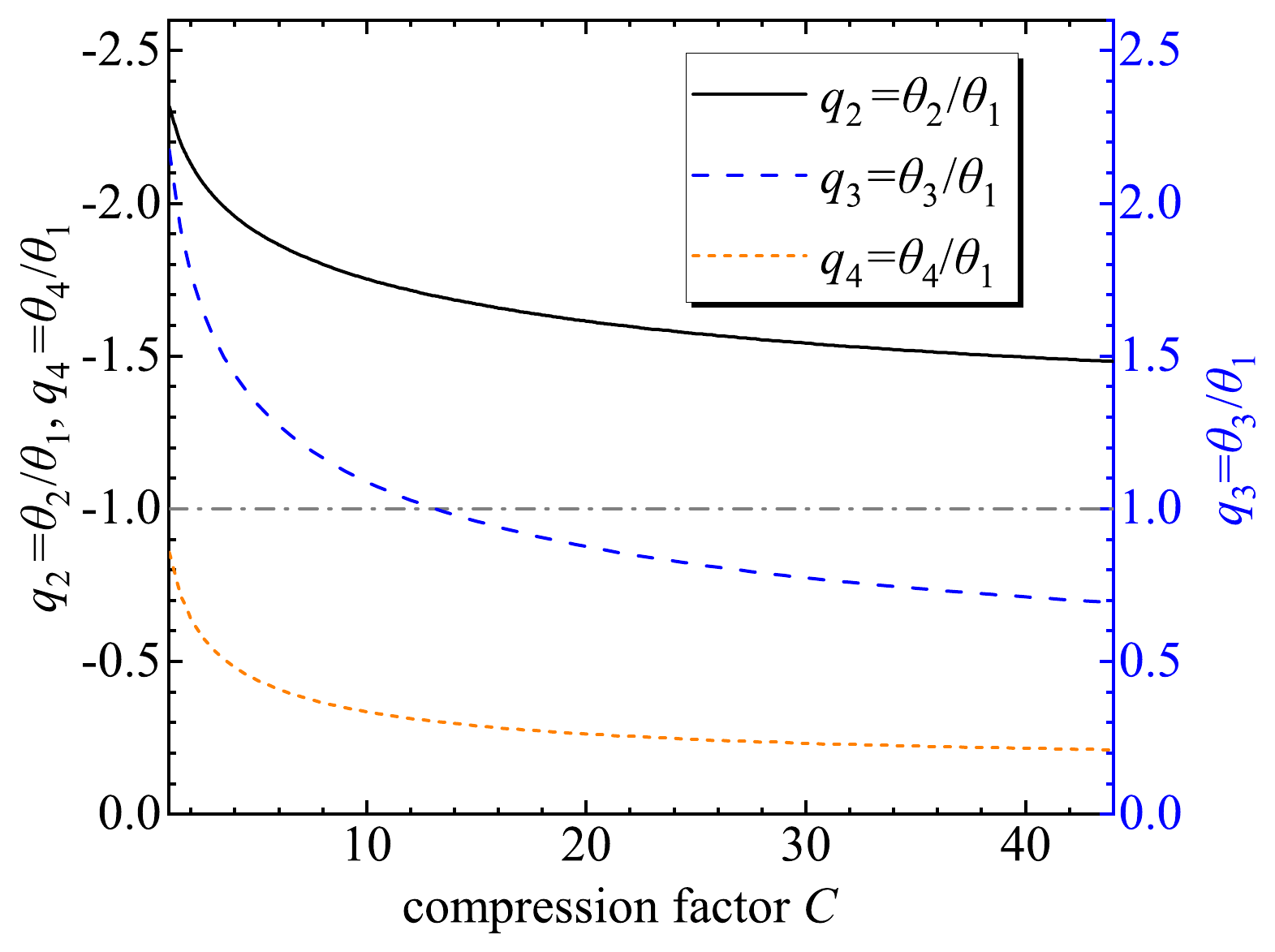}
	\caption{The solved  $q_2=\theta_2/\theta_1,q_3=\theta_3/\theta_1$ from Eqs. \eqref{eq:condition1}, \eqref{eq:condition-position}  in terms of   compression factor $C$ for Case 3. The  corresponding $q_4=\theta_4/\theta_1$ is also included. The black dashed curve is plotted for  clear  comparison with $q_2, q_3, q_4$.  }
	\label{fig:s-chicane-0.25} 
\end{figure}

\begin{figure}[h]
	\centering
	\includegraphics[scale=0.33]{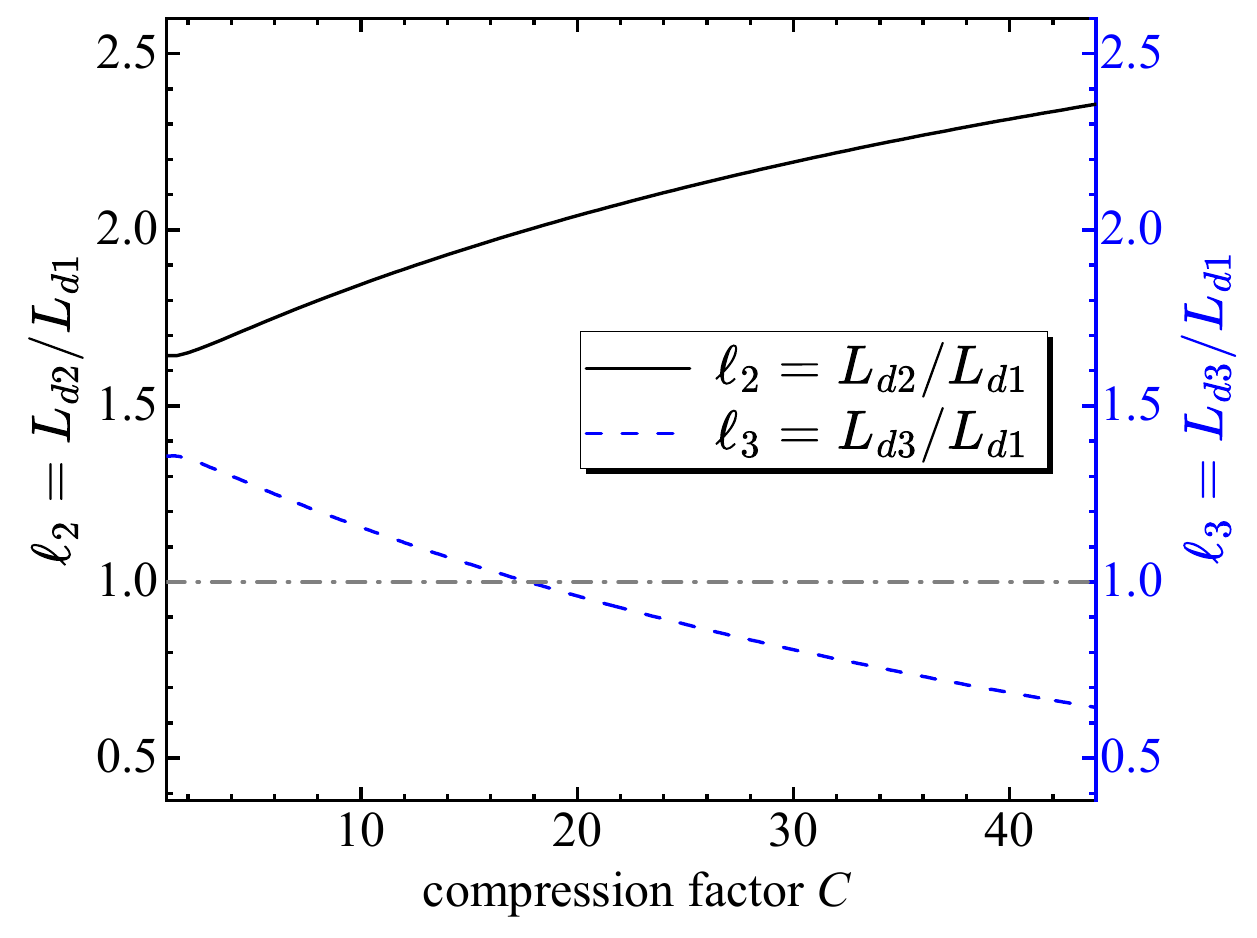}
	\caption{ The corresponding
		 $\ell_2=L_{d2}/L_{d1}, \ell_3=L_{d3}/L_{d1}$ after solving the  $q_2=\theta_2/\theta_1,$ and $ q_3=\theta_3/\theta_1$ for Case 3. The black dashed curve is plotted for  clear  comparison with $\ell_2$ and $\ell_3$.  }
	\label{fig:s-chicane-0.25-LD} 
\end{figure}

\textit{For Case 3},  the   bending radii are set to $    \rho_1=-\rho_2  =\rho_3=-\rho_4$.   
%\begin{equation}  \label{eq:condition5}
%	\frac{\theta_1\theta_2L_{d1}}{\theta_3\theta_4L_{d3}}=\frac{k_3 \theta_3  -k_4 \theta_4   }{k_1 \theta_1 -k_2 \theta_2}
%\end{equation}
%and
%\begin{equation}   \label{eq:condition6}
%\frac{	L_{d2} }{L_{d1}}= -\frac{\theta_1}{\theta_3} \frac{    \theta_3(\delta_3-\delta_1)+(\theta_2+\theta_3) \delta_2 }{\theta_1  \delta_2  +(\theta_1+\theta_2)(\delta_3-\delta_1)}  
%\end{equation}  
Given  the complexity of the obtained   CSR cancelation conditions  in Eq. \eqref{eq:condition1} (see Appendix \ref{appendixki}), % cannot be solved to obtain  analytical expressions for $q_2=\theta_2/\theta_1, q_3=\theta_3/\theta_1$ as a function of   factor $C$, 
we calculate the results of  $q_2=\theta_2/\theta_1,q_3=\theta_3/\theta_1$   in terms of the compression factor $C$,   as  shown in Fig. \ref{fig:s-chicane-0.25}.  
%  This result can be checked by solving Eqs.   \eqref{eq:condition1} and \eqref{eq:condition-position} under the conditions in 	Eq. \eqref{eq:case-1-2}.  
Using  the obtained $q_2 $ and $ q_3 $, 
we also compute  the values of $q_4, \ell_2, \ell_3$ from Eq. \eqref{eq:condition-position} as shown in Figs.  \ref{fig:s-chicane-0.25} and \ref{fig:s-chicane-0.25-LD}, which   varies only with compression factor $C$. 
%Finally,  for a concrete compression goal, one can obtain the bending angle  and drift relation for a chicane in their entirety with the aid of the provided Figs.  \ref{fig:s-chicane-0.25} and \ref{fig:s-chicane-0.25-LD}. 

\begin{figure}[h]
	\centering
	\includegraphics[scale=0.27]{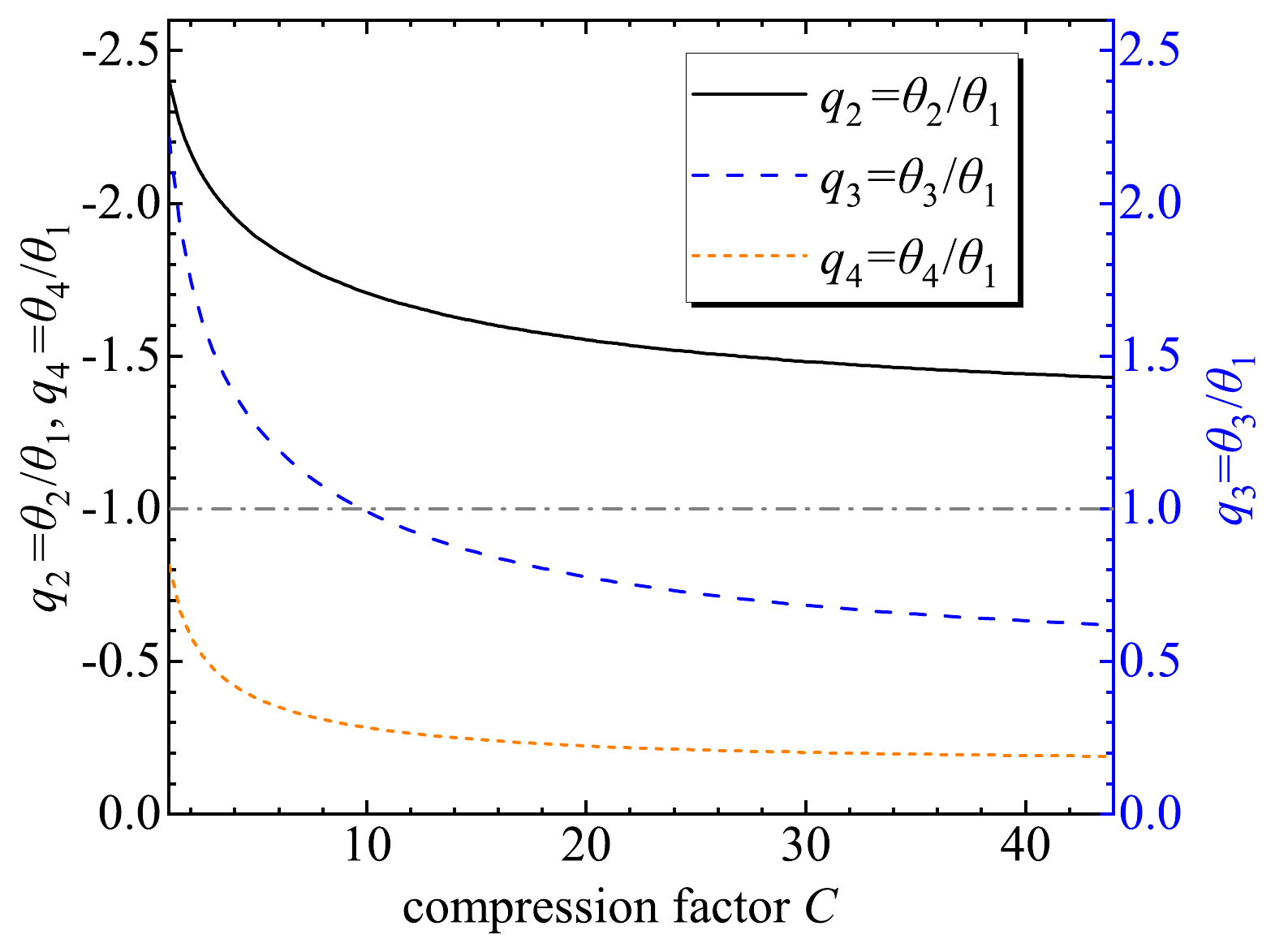}
	\caption{The solved  $q_2=\theta_2/\theta_1,q_3=\theta_3/\theta_1$ from Eqs. \eqref{eq:condition1}, \eqref{eq:condition-position}  in terms of   compression factor $C$ for Case 4. The  corresponding $q_4=\theta_4/\theta_1$ is also included. The black dashed curve is plotted for  clear  comparison with $q_2, q_3, q_4$.   }
	\label{fig:s-chicane-0.25-sameLB} 
\end{figure}

\begin{figure}[h]
	\centering
	\includegraphics[scale=0.33]{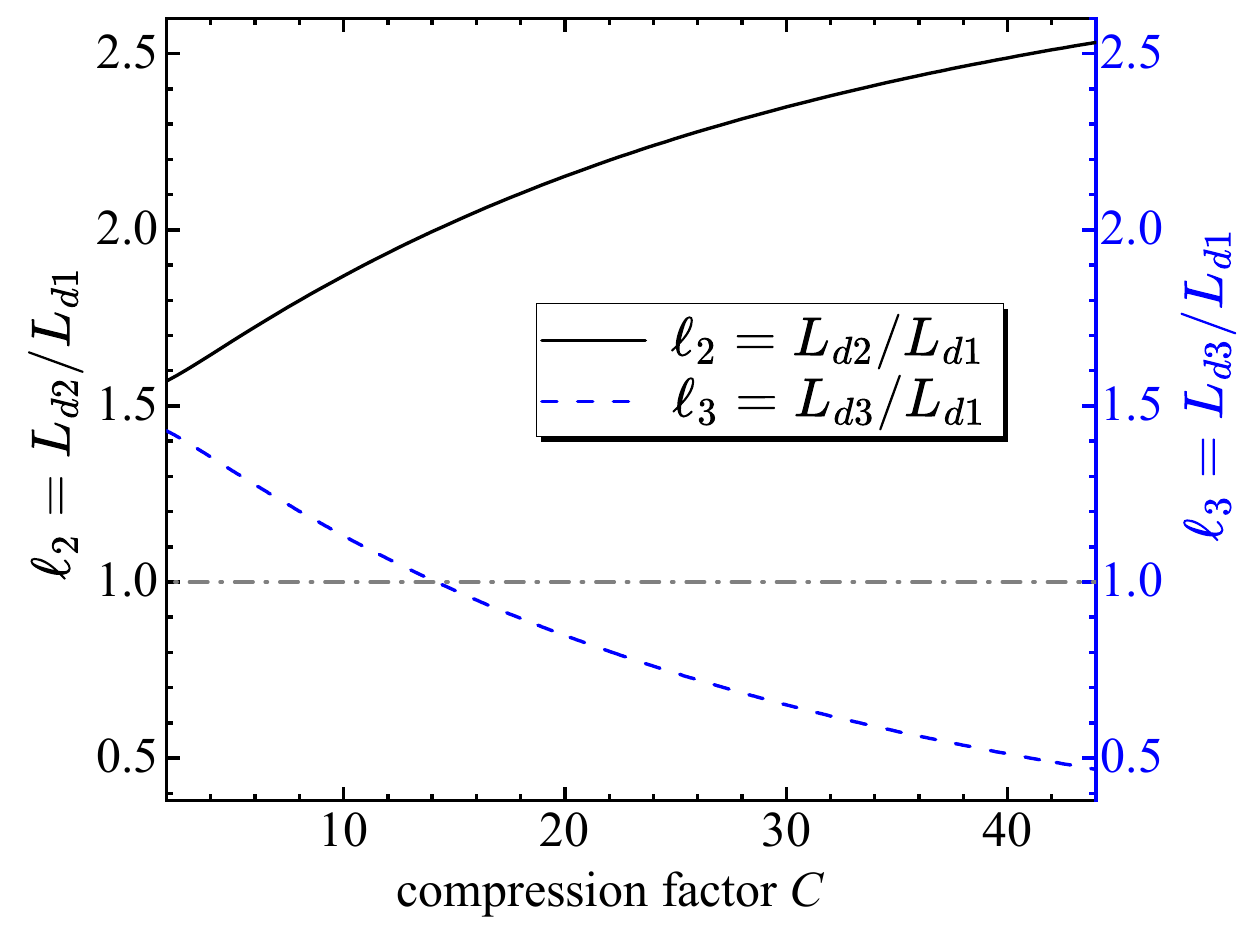}
	\caption{  The corresponding
		$\ell_2=L_{d2}/L_{d1}, \ell_3=L_{d3}/L_{d1}$ after solving the  $q_2=\theta_2/\theta_1,$ and $ q_3=\theta_3/\theta_1$ for Case 4.  The black dashed curve is plotted for  clear  comparison with $\ell_2$ and $\ell_3$.    }
	\label{fig:s-chicane-0.25-LD-sameLB} 
\end{figure}

\textit{For Case 4}, the dipole lengths are fixed as $ L_{B1}=L_{B2}=L_{B3}= L_{B4}$. %Thus the bending radii ratios  can be determined as $\rho_2/ \rho_1=1/q_2 $,  $\rho_3/ \rho_1=1/q_3 $, $\rho_4/ \rho_1=-1/(1+q_2+q_3)$.  
 In a similar manner to Case 3, the  values of $q_2, q_3 $  can be solved   from Eqs. \eqref{eq:condition1} and  \eqref{eq:condition-position} under the conditions in 	Eq. \eqref{eq:case-1-2}.  Correspondingly,   the  quantities $q_4, \ell_2, \ell_3$ for Case 4 are plotted in Figs. \ref{fig:s-chicane-0.25-sameLB} and \ref{fig:s-chicane-0.25-LD-sameLB}. 
We  observe that the results of $q_2 ,q_3, q_4, \ell_2, \ell_3$ in Case 4 shows great similarity with those of Case 3.
% and with a greater speed decrease than in Case 3.

We find that, the theoretical results for both Case 3 and Case 4  indicate that the last two dipoles are weaker (shorter) than the first two dipoles for the asymmetric S-chicane.   More specifically, the 3rd (4th) dipole is shorter   or  weaker than the 2nd (1st) dipole for   Case 3 or Case 4, respectively.  This weakening or shortening effect becomes more pronounced as the compression factor increases. 
We also find that a  symmetric S-chicane fails to  satisfy the  CSR cancelation conditions in our calculation, as none of the compression factors  $C$ in Figs. \ref{fig:s-chicane-0.25} and \ref{fig:s-chicane-0.25-sameLB}  meet the requirement that both $q_2 =-2$ and $q_3=2$.  %Then an  interesting  question arises, whether a CSR-cancelled chiacne  can be achieved, by fine adjustments of a symmetric S-chicane, comprising of four equal-length dipoles with fixed dipole positions. To this end,  just adjust the strength of the dipoles to $q_2=-1.62,   q_3=0.87, q_4=-0.25$ from Fig. \ref{fig:s-chicane-0.25-sameLB} (or Eq. \eqref{eq:condition1}).  Because the compression factor $C$ is equal to $14.2$ in case 2 when  $L_{d2}/L_{d1}=2, L_{d3}/L_{d1}=1$ is satisfied from  Fig. \ref{fig:s-chicane-0.25-LD-sameLB}. Finally, the CSR-immune   S-chicane  is achieved.  Of course,   further simulations and even experimental verification will be essential. 

 \subsection{ Numerical verification   of the proposed conditions }\label{mo:S}

\begin{figure}[hb] 
 	\includegraphics[scale=0.32]{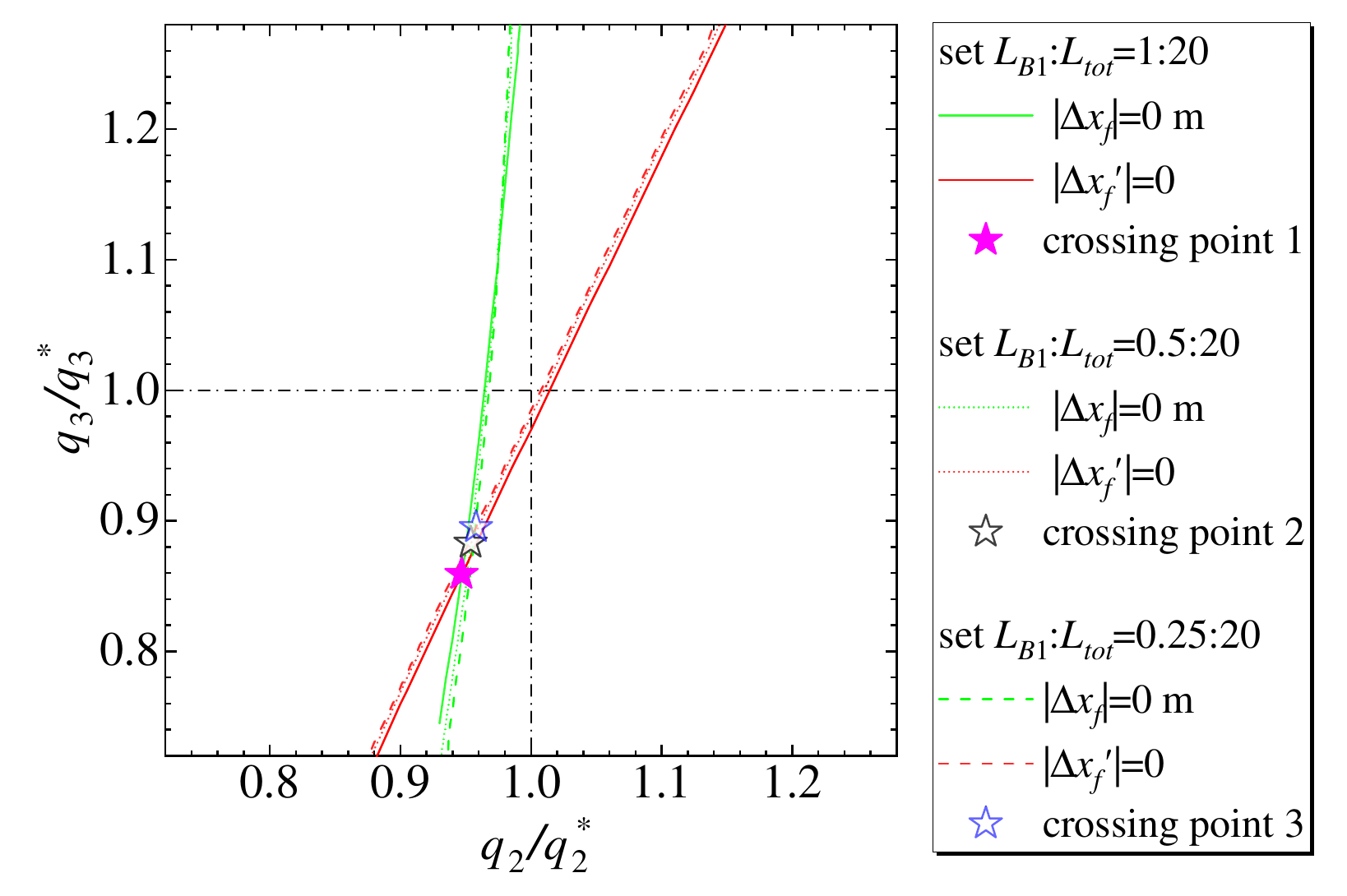}
 	\caption{The ($q_2/q_2^*, q_3/q_3^*$) of a  chicane  with $|\Delta x_f|=0$ m and $|\Delta x_f^\prime|=0$, as shown in the green and red   solid   curves. 
  As a comparison, the result for $L_{B1}:L_{d1}$=1:20, 0.5:20 and 0.25:20 are calculated  as  $(  q_2/q_2^*,  q_3/q_3^*) =$(0.947, 0.859), (0.954, 0.883), and   (0.958, 0.895), respectively. %The curves and points have the same meaning as in Fig. \ref{fig:cross}. 
 	}  
 	\label{fig:midu222} 
 \end{figure}

 To investigate the impact of the assumptions, including the constant bunch length  in one dipole  and the neglected dipole length,  we perform numerical integration and ELEGANT   analysis for the asymmetric S-chicane.
To begin with, we  present the results of the numerical integration. 
  First, the dipole lengths are taken into account.  The expressions of $\ell_2, \ell_3$ in Eq. \eqref{eq:ell-2+3} and the  $	R_{56}^{s_0\to s_f}$, $L_{tot}$ are recalculated, the details of which can be found in Appendix  \ref{r56-tot}.  
Second, we consider a gradually varying bunch length   in a similar manner to the asymmetric C-chicane using an integration method    \cite{Emma:1997hj}. These variables $R_{16}^{s \rightarrow s_f}$ and $R_{26}^{s \rightarrow s_f}$  in Eq. \eqref{eq:integrate}, and $R_{56}^{s_0\to s}$ in $k_1,k_2,k_3,k_4$,     change  with the position $s$, as expressed in Appendix \ref{appendix1}.   
Eventually, the  CSR cancelation conditions can be obtained   and  compared with the results  $q_2^* =-1.75$, $q_3^* =1.09$ (Eq. \eqref{eq:condition1})  for  Case 3 when the  compression  factor  is $C=10$.  
   Figure  \ref{fig:midu222}  (pink pentacle) shows the results from integration method as $(q_2/q_2^*,q_3/q_3^*) =(0.947, 0.859)$,  indicating  a relatively accurate calculation of the point-kick model.     
 Similar to the asymmetric  C-chicane,  the  $(q_2, q_3 )$ obtained by integration method  is  closer to the result of the point-kick model  after reducing the ratio $L_{B1}:L_{tot}$, as shown in  Fig. \ref{fig:midu222}. 
 %To some extent, the model calculation trade this slight loss in accuracy  for  broader CSR cancelation.  
  
\begin{figure}[ht]  
 	\centering
 	\includegraphics[scale=0.22]{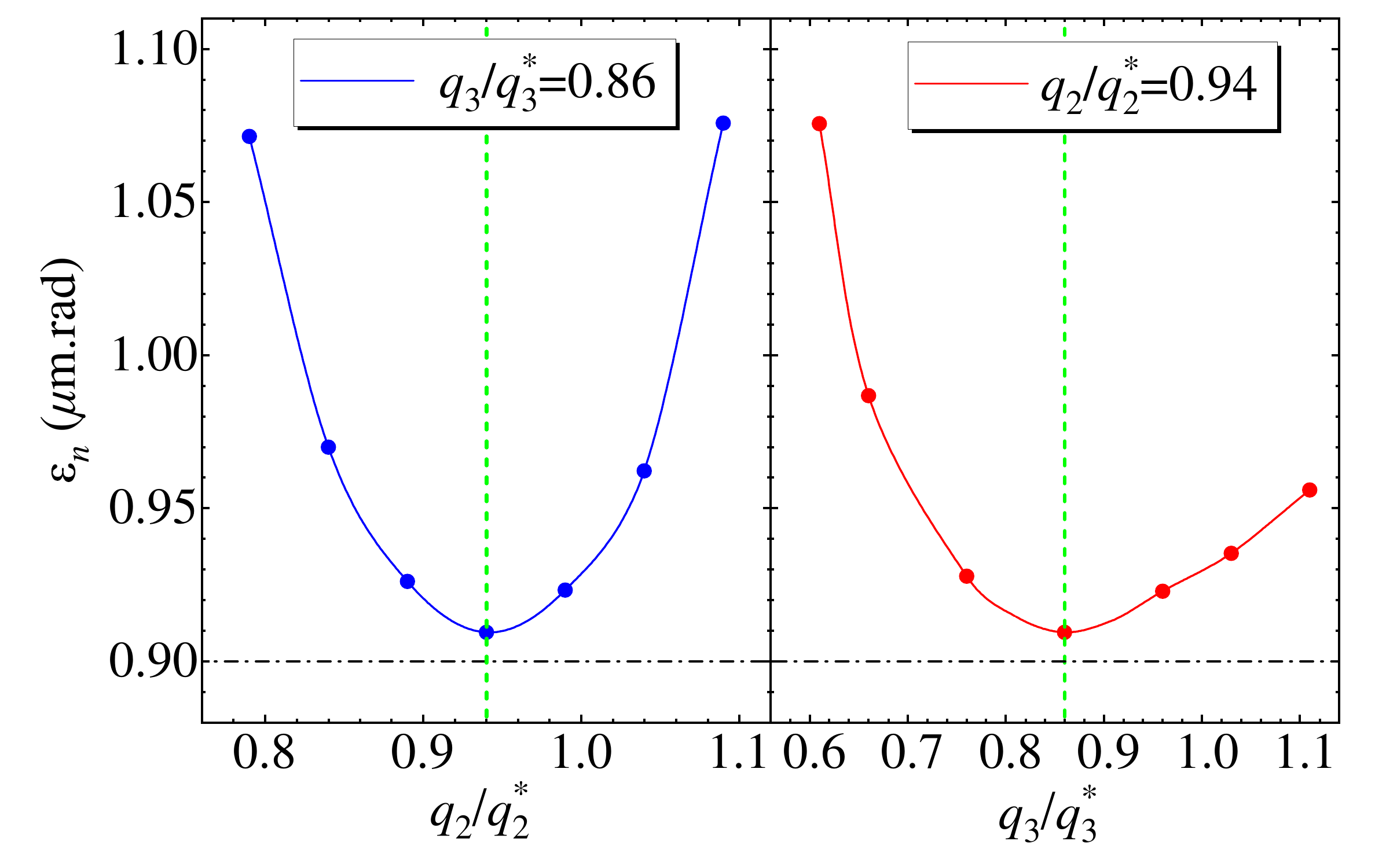} 
 	\caption{The ELEGANT simulations result of the  CSR-induced emittance $\varepsilon_n$ near $(q_2^*,q_3^*)$ for    asymmetric S-chicane.  The simulations results  turn  out that the parameters satisfy   $(q_2/q_2^*, q_3/q_3^*)=( 0.94, 0.86)$   with scaned initial Twiss parameters $(\alpha_{x0},\beta_{x0})= ( 1.5, 32 \text{ m})$, as shown in the dashed curves. }
 	\label{fig:S-chicane-elegant} 
 \end{figure}

Furthermore, the minimum transverse  emittance  $   \varepsilon_n$ for the asymmetric S-chicane, together with the corresponding $(q_2, q_3)$, is searched via ELEGANT simulations. 
The scaned results $(q_2/q_2^*, q_3/q_3^*)=( 0.94, 0.86)$ are obtained. 
 For illustration,   the  normalized emittance growth near $(q_2/q_2^*, q_3/q_3^*)$ is displayed in Fig. \ref{fig:S-chicane-elegant}. Correspondly, 
the complete  asymmetric S-chicane information can be found in  Table \ref{Table-S-chicane}. Our results from point-kick analysis are consistent  with these numerical results basically.

%%%0.947 0.896  $q_2^* =-1.71$, $q_3^* =0.99$   for case 2  $(\Delta q_2/q_2^*,\Delta q_3/q_3^*) =(0.8\%, 2.1\%)$. 

% Various similar numerical integrations have also been conducted for different compression factor $C$. 
%One main reason is that the specific value of $L_{B1}$ can result in a meager influence. For example,  a comparison  of the $(l,q)$ value under different dipole lengths is carried out in Fig. \ref{fig:midu222}. Here the present result of $L_{B1}=0.5$ m is compared with the result of   $L_{B1}=1$ m. As shown in Fig. \ref{fig:cross}, a normalization error   is calculated as  $(\Delta q_2/q_2^*,\Delta q_3/q_3^*) =(0.7\%, 2.4\%)$. 

%Lastly, the $(q_2/q_2^*,q_3/q_3^*)$ of a   chicane for different compression factor under case 1 and case 2 are plotted in Fig. \ref{fig:modify-S-case1}.  The  modified  calculation  results of  $(q_2/q_2^*,q_3/q_3^*)$   change  slightly with the factor $C$, and  basically in the  range (0.9,1), which shows a relatively accurate calculation by the point kick model. This modification is meaningful because it not only applies to any $R_{56}^{s_0\to s}$, but  has a slight influence on $L_{B1}:L_{tot}$, as the example in Fig. \ref{fig:midu222}  shows.

\begin{table}
	\caption{Parameters of the symmetric and  asymmetric  S-chicane settings for Case 3 with   units in meters.} \centering
	\begin{tabular}{lcccc}
		\hline	\hline\noalign{\smallskip}
		& \multirow{2}*{  Symbol }  &  symmetric	     &  asymmetric	 	   \\  \noalign{\smallskip}
		&     &   	S-chicane	    &   	S-chicane	   &   \\ \hline		\noalign{\smallskip}
	Bending radii of each dipole &$ \rho$   &20.66 & 14.46 &     \\\noalign{\smallskip}
%	Length of the first   dipole ~~~ &$ L_{B1} $   & 0.50 &    m  \\  
Length of the 2nd  dipole ~~~ &$  L_{B2}$  &2.0 & 1.65 &       \\  \noalign{\smallskip}
Length of the 3rd dipole  &$ L_{B3} $  &2.0 &  0.94 &       \\  \noalign{\smallskip}
Length of the 4th dipole  &$ L_{B4}$  &1.0 &  0.29 &       \\  \noalign{\smallskip}
Length of the 1st drift &$ L_{d1}$   &3.25 & 3.18 &    \\  \noalign{\smallskip}
Length of the  2nd drift &$ L_{d2} $ &7.5  & 8.10 &    \\  \noalign{\smallskip}
Length of the 3rd drift &$ L_{d3}$   &3.25  & 4.85 &     \\   
		\hline	\hline
	\end{tabular} \label{Table-S-chicane}
\end{table}

\section{comparation among  symmetric and asymmetric C- and S-chicanes}\label{sec.com}
In this section, to further demonstrate the CSR suppression efficiency of our proposed asymmetric chicane designs, we compare the performance in suppressing the CSR-induced emittance growth among the four types of chicanes, including symmetric  C- and S-chicanes, and the   asymmetric   C- and S-chicanes.  These chicanes have set the same  $L_{tot}$, $	R_{56}^{s_0\to s_f}$, $L_{B1}$, and compression factor $C$,  as summarized in Table   \ref{Table-same-Parameter}. Whereby one can obtain the   other chicane parameters of  symmetric  S-chicane, as presented in  Table  \ref{Table-S-chicane}. 
 These chicanes are tested  by ELEGANT  with the initial bunch parameters in Table  \ref{Table:bunch}.

By scanning   the initial C-S parameters, 
we aim to achieve minimum emittance growth at the exit of the symmetric   C- and S-chicanes.
By doing this, the  simulation results for every chicanes can be obtained by ELEGANT and can be seen in Table \ref{Table-C}.
 In the case of the asymmetric C- and S-chicanes,  the normalized emittance growth  $\Delta \varepsilon_{n}/\varepsilon_{n0} $ is reduced by more than tenfold  compared with  the symmetric ones,   It appears  that the emittance growth due to CSR can be effectively suppressed. 

In the above sections, the  emittance growths  induced by dipole-CSR have been verified for  asymmetric C- and S-chicanes. However,  some study has shown that a longer drift space between the dipoles  may degrade the emittance significantly, as  the CSR effects dominate in the drift space   \cite{Borland:2001xv}.  
  Therefore,   it is worthwhile to investigate more detail CSR effects  (denoted by all-CSR), considering the transient CSR at the dipole edges, incoherent synchrotron radiation in the bending magnets, the classical, single-particle synchrotron radiation, and the CSR effect in the dipoles and the following drift space with the Stupakov model \cite{Borland:2000gvh}.  
%In the presence of these effects, we find that the increase in emittance is mainly due to the  CSR effects in the drift space between the dipoles, which is consistent with the conclusion in  \cite{Borland:2000gvh}.  
Furthermore,  we  simulate the emittance growth of the symmetric (asymmetric) C- and S-chicanes, taking into account the all-CSR effects. The results are listed in Table \ref{Table-C}. 
We find that the relative emittance growth of the symmetric C- and S-chicanes  are 5.5 and 8.9 times larger than that of asymmetric ones, respectively. 
Overall,  the asymmetric   C- and S-chicanes have good success in  suppressing of the  emittance growth, whether only the dipole-induced CSR or the all-CSR is considered. 
 %Therefore, it is essential to investigate the CSR suppression efficiency using a more detailed physical model of the CSR wake in order to reduce this effect by appropriately deviating from the current result.

 \begin{table}
 	\caption{Results of the ELEGANT simulations of the  finally  geometric emittance $\varepsilon_{nf} $, and relative emittance growth $\Delta \varepsilon_{n}/\varepsilon_{n0}  $  for symmetric and asymmetric C- and S-chicane with initial  emittance $\varepsilon_{n0} =0.9$ $\mu$m.rad.} \centering 
 	\begin{tabular}{lcc}
 		\hline	\hline\noalign{\smallskip} 
 		~~~CSR effects   & ~~~ $\varepsilon_{nf} $ ($\mu$m.rad) ~~~  & ~~~ $\Delta \varepsilon_{n}/\varepsilon_{n0}  $ ~~~ \\\hline \noalign{\smallskip} 
 		\multicolumn{3}{c}{  asymmetric C-chicane }	 \\\hline \noalign{\smallskip} 
 		~~~steady-state CSR		 &0.905& 0.53\%  \\\noalign{\smallskip} 
 		~~~all-CSR		 &0.962 &6.86\%  \\
 		 	\hline\noalign{\smallskip} 
 		\multicolumn{3}{c}{ symmetric C-chicane } \\\hline \noalign{\smallskip} 
 		~~~steady-state CSR	  & 1.0  &11.0\% \\\noalign{\smallskip} 
 		~~~all-CSR	  &1.238  &37.6\%  \\\noalign{\smallskip} 
 	\hline \noalign{\smallskip} 
		\multicolumn{3}{c}{  asymmetric S-chicane }	 \\\hline \noalign{\smallskip}
	~~~steady-state CSR  &   0.910 & 1.1\%   \\\noalign{\smallskip}
	~~~	transient CSR  &    0.950  &5.55\%   \\ 
		\hline\noalign{\smallskip}
		\multicolumn{3}{c}{ symmetric S-chicane } \\\hline \noalign{\smallskip}
	~~~steady-state CSR & 1.096 & 21.8\% \\\noalign{\smallskip}
	~~~transient CSR 	  &1.344 & 49.3\%  \\ 
	\hline	\hline \noalign{\smallskip} 
 	\end{tabular} \label{Table-C}
 \end{table}

\section{SUMMARY AND DISCUSSION}\label{summary}

This paper reports an analytical expression of the CSR net kick, thus finds a solution to cancel the CSR-induced transverse emittance  growth in a chicane. The theoretical CSR cancellation conditions, derived from the point-kick model,  indicate an asymmetric chicane design for both C and S chicane  geometries. 
%  theoretical designs for asymmetric  $C$- and S-chicane bunch compressors, both of which can suppress the CSR-induced transverse emittances during bunch compression. 
The design of the asymmetric C-chicane  may be valuable for   improving on the current numerous symmetric C-chicanes.  Specifically,  it is required to adjust the bending  angle  ratio between the first two dipoles and the last two, as well as the length ratio between the 1st and 3rd  drifts. Despite the presence of a ``negative drift''   between the 2nd and 3rd dipoles, this can be realized by adding quadruples in 2nd drift section.  Correspondingly, the ratio  $q_3$ and the effective value of the ``negative drift'' can be approximated to be a simpler expression as Eqs. \eqref{eq:q3l2}   and \eqref{eq:q4}. 
For the asymmetric S-chicane, the 1st and 3rd (2nd and 4th) dipoles have the same bending direction. The chicane information, including   the bending angles  and draft legnth ratios,  are shown in  Figs.   \ref{fig:s-chicane-0.25},  \ref{fig:s-chicane-0.25-LD} and \ref{fig:s-chicane-0.25-sameLB},  \ref{fig:s-chicane-0.25-LD-sameLB}   for the cases of fixed bending radii and fixed dipole lengths, respectively. 
 %the value of $(\theta_2/\theta_1, \theta_3/\theta_1)$ can be solved by solving  equations, including the position condition in Eq. \eqref{eq:condition-position},  and the CSR cancelation conditions  in Eq. \eqref{eq:condition1}  for the cases of fixed bending radii and fixed dipole lengths, respectively. 
Roughly, both the values of $(\theta_3/\theta_1, L_{d2}^{\text{eff.}}/L_{d1})$ for C-chicane and $(\theta_2/\theta_1, \theta_3/\theta_1)$ for S-chicane  can be determined as  compression factor $C$-dependent quantities.
Such a property  enables our results easily transferable to arbitrary chicane design. 
  %, and the value of $\theta_1$ and $L_{d1}$  depends entirely on the design goal of the chicane, namely the value of $L_{tot}$ and $	R_{56}^{s_0\to s_f}$. 

As a verification of these theoretical results,  numerical  integration calculations and  ELEGANT simulations for asymmetric C- and S-chicanes are presented, which show a slight shift with respect to the theoretical result.   However our result remains valuable as it reveals the main picture of cancelling the CSR-driven emittance excitation. For an actual design,  the theoretical results can be considered as  initial values and then the design can be optimized by fine-tuning the chicane parameters. 
 Moreover, compared with the symmetric C- and S-chicanes with the same  compression target,   the proposed asymmetric chicanes, show a promising performance in suppressing the emittance growth in our simulations.  
%$	L_{tot},	R_{56}^{s_0\to s_f},  L_{B1}$ and

One can find that the key to supressing the CSR effect is to weaken the strengths (or shorter the lengths) of the last two dipoles according to certain rules,    for both C- and S-chicanes.  Such weakening (shortening) becomes more pronounced as the compression factor increases. Note that the above features coinside  with the asymmetric DBA-based bunch compressor introduced in  \cite{Zhang:2023cgl}, where the last dipole is $C^{-1/3}$ or $C^{-1/2}$ times weaker than the 1st dipole.  An intriguing pattern emerges that the  asymmetric C- and S-chicane, as well as the  DBA-based bunch compressor,   share similarities and delicately suppress the CSR-induced emittances growth during the compression process.  
 
While the primary focus of this paper lies in analyzing the design principles that effectively suppress CSR-induced emittance growth in the chicane bunch compressors, we also assess the potential  microbunching instability associated with the lattice and the beam parameters. Using our developed semi-analytical Vlasov
solver \cite{Tsai:2017hef,Tsai:2020ddc} enables the fast, efficient evaluation of the various  lattice designs. The semi-analytical calculations indicate that
these chicane designs, which are effective in suppressing CSR-induced emittance growth, also exhibit a well-controlled MBI. Specifically, based on the beam parameters and chicane settings presented in Tables  
\ref{Table-same-Parameter}, \ref{Table-C-chicane},   \ref{Table:bunch}, and \ref{Table-S-chicane}, 
taking into account both the steady-state and transient CSR and the
longitudinal space charge effects, the simulation results
for the  asymmetric C-chicane reveal a maximum gain of about 3.5 occurring
around an initial modulation wavelength of 50 $\mu$m. In the case of the  asymmetric S-chicane, simulation results demonstrate a maximum 
gain of 4 around a similar modulation wavelength of 50 $\mu$m. 
It is worth mentioning that this paper does not consider other collective effects,  such as  space-charge forces,  linac geometric wake field, and many others. For a physically realistic scenario,  these are difficult to cancel out and need to be fully taken into account.

We hope that our results can be used as a starting point for practical design, construction and experimental studies on the CSR-immune four-bend chicane compressors. 
A feasible verification scheme based on the current symmetric C-chicane is, to insert a ``negative drift'' section between the 2nd and 3rd dipoles, and to increase the strength of the first two dipoles and  decrease the last two, while adjusting the longitudinal position of the 2nd and 3rd dipoles by unity. 
On the other hand, the confirmation of the asymmetric S-chicane design is easily achievable.  Based on a symmetric S-chicane,   move the position of the 3rd dipole and vary the current of the dipoles to control their bending strength.  In addition, asymmetric S-chicane design is also available based on the  currently widely used C-chicanes. To obtain the asymmetric S geometries, changing the positive and negative poles of the last two dipole  currents seems to be a convenient approach. 
  This paper opens up new avenues for the realisation of CSR-immune bunch compressor, and provide a good guidance for accelerator scientists to respond to the future development of chicane bunch compressor.   
  
%we aim to propel accelerator science forward and contribute to the development of cutting-edge accelerators with enhanced performance and versatility for scientific and industrial applications

\section*{ACKNOWLEDGMENTS}  
  This work was supported by the National Natural Science Foundation of
China (No. 12275284  and No. 12275094), and the Fundamental Research Funds for the Central Universities (HUST) under Project No. 2021GCRC006.
We thank  Cai Meng and Wei Li of IHEP for useful discussions.

%xiangdao-----In conclusion, we have demonstrated a novel bunch compression technique that allows us to achieve 50 fs resolution in MeV UED, opening new opportunities for studies of dynamics on sub-100 fs timescale.   This technique should be easily transferable to other facilities In  addition to MeV UED, ultrashort and ultrastable electron beam is also essential for external injection in plasma acceleration, THz-driven acceleration, inverse Compton scattering x-ray source, etc. We expect this technique to  have a strong impact in future development of ultrashort electron beam based scientific facilities and applications

%We have proved that four is the minimum number of dipoles required to cancel the CSR, therefore, more design scheme based our asymmetric S-chicane is realizable, such as   five -bend  and  six-bend by separating the second and third dipoles into  !

 \begin{appendices}

\section{$k_i$ for different dipoles}\label{appendixki}
 	
Here we assume that the bunch length is constant in a single dipole for point-kick model, so each magnet corresponds to a constant $k_i$. Concretely, the $k_1$ for the 1st dipole is $k_1 = k_0=	 0.2459\frac{N_b r_e}{\gamma \sigma_{z0}^{4/3}}  $, reflecting the  bunch length at the chicane entrance;  the $k_2$ and $k_3$ for the 2nd and 3rd dipoles reflect  the bunch lengths when crossing the middle position of the dipoles; and the last $k_4 $ reflects the  bunch length at the chicane exit.   
 %	Actually, $k_i$ changes in every position of the dipoles, and this scenario will be considerd in the revision subsection. 
 	 
The value of $k_2$ for the 2nd dipole is related to  $k_0$ as
\begin{equation}
k_2=k_0\frac{\sigma_{z2}^{-4/3} }{\sigma_{z0}^{-4/3}}=\frac{k_0}{(1+h~ R_{56}^{s_0\to 2\text{m}})^{ 4/3}  } ,
\end{equation}
 because the  bunch length  $\sigma_{z} \approx \sigma_{z0}(1+h R_{56}^{s_0\to s}) $ with 

\begin{equation}
h=\frac{1-C}{C~ R_{56}^{s_0\to s_f}}, ~~~ R_{56}^{s_0\to s_f}=  -  L_{d1} \theta_1^2 \left(  1 +  \ell_{3}   q_3^2   \right),
\end{equation}
 here $ C=\sigma_{z0}/\sigma_{zf}$ is the bunch compression factor,   $ h$ is the linear chirp  related to RF cavity phase,  $R_{56}^{s_0\to s}$ is the   first-order longitudinal momentum compaction at lacation $s$ and the notation “$ s_0\to 2\text{m}$"  in  $R_{56}^{s_0\to 2\text{m}}$ means transport from  the chicane entrance to  the midpoint of the second magnet, which can be written as
 	
 	\begin{equation}
 R_{56}^{s_0\to 2\text{m}} =-\frac{ \theta_1^2}{2}   L_{d1} .  
\end{equation}
 	The value of   $k_3$ for the 3rd dipole is similar to the expression of $k_2$ as
 	
 	\begin{equation}
k_3=k_0\frac{\sigma_{z3}^{-4/3} }{\sigma_{z0}^{-4/3}} =\frac{k_0}{(1+h~ R_{56}^{s_0\to 3\text{m}})^{ 4/3}  } ,
\end{equation}
 	with 
 	\begin{equation}
 R_{56}^{s_0\to 3\text{m}} =-  L_{d1}   \theta_1^2 (1+  \frac{1}{2}   \ell_3   q_3^2  ) .\end{equation}
 	The last $k_4 $ reflects the  bunch length at the chicane exit   (denoted by subscript “$s_f$") as
 	
 	\begin{equation}
k_4=k_0\frac{\sigma_{z4 }^{-4/3} }{\sigma_{z0}^{-4/3}} =C^{4/3} k_0  .
\end{equation}
 	
 Equations \eqref{eq:q2},	\eqref{eq:simpm12},
  \eqref{eq:q4} are the result that applying the $k_i$ under the assumption that
  the length of $L_{B} $ is much smaller than  $L_{d}$.
 	
For S-chicane,	the  $k_1$ and $k_4$   have the same expression as the C-chicane. 
 	And the   difference is reflacted by $k_2$ and $k_3$ as
  	\begin{equation}
k_2=k_0\frac{\sigma_{z2}^{-4/3} }{\sigma_{z0}^{-4/3}}=\frac{k_0}{(1+h~ R_{56}^{s_0\to 2\text{m}})^{ 4/3}  } , 
\end{equation}
 	with 
\begin{equation}
	\begin{aligned}
		h=\frac{1-C}{C~ R_{56}^{s_0\to s_f}}&, ~~~ R_{56}^{s_0\to s_f}=    \theta_1^2  L_{d1} ( q_2+q_3+ \ell_{2} (1+q_2)  q_3    ),   \\
		&R_{56}^{s_0\to 2\text{m}} = \frac{ L_{d1}   \theta_1^2  q_2  }{2}.
	\end{aligned}
\end{equation}
 		 
 		The value of   $k_3$ for the 3rd dipole can be written as 
  	\begin{equation}
k_3=k_0\frac{\sigma_{z3}^{-4/3} }{\sigma_{z0}^{-4/3}}=\frac{k_0}{ (1+h~ R_{56}^{s_0\to 3\text{m}})^{ 4/3}}  ,
\end{equation}
with 
\begin{equation}
	R_{56}^{s_0\to 3\text{m}} =\frac{1}{2} L_{d1} \theta_1^2  \left( \ell_2 (  1+ q_2) q_3+  ( 2q_2+ q_3)   \right)    .
\end{equation}

According to the above calculations, it is not hard to find that the change of bunch length in chicane during the compression process mainly occurs in the 2nd and 3rd dipoles,  regardless of C- or S-chicane.

 	\section{The results considering the dipole lengths}\label{r56-tot}
For asymmetric C-chicane:

After considering the dipole lengths, the drift length between the last two dipoles  can be obtain according to Eq. \eqref{eq:achromatic-condition-all} as
 \begin{equation}
\ell_{3}=-\frac{1 }{q_3}-\frac{ L_{B1} }{q_3 L_{d1}}  -\frac{L_{B3}}{L_{d1}}.
\end{equation}
 The    $	R_{56}^{s_0\to s_f}$, and  $L_{tot}$  can be written as 
 \begin{equation}
 	\begin{aligned}
  	L_{tot}=&L_{d1}(1- \frac{1}{q_3})-L_{B1}(\frac{1}{q_3}-2  ) + L_{B3} +L_{d2}^{\text{real}},\\
 R_{56}^{s_0\to s_f}=&	-\frac{\theta_1^2}{3} \left( 3L_{d1} (1-q_3)  +L_{B1}( 2-3q_3 )-q_3^2 L_{B3}  \right).
 	\end{aligned}
 	\end{equation}
  Here we approximate that  the total length are the sum of  $L_{d1}, L_{nd}, L_{d3}$  and four dipole lengths.  
      Thus   $L_{d1}$ and   $\theta_1$ can be expressed as
   \begin{equation}
  	\begin{aligned}
  		L_{d1}=&\frac{ q_3(L_{tot}-L_{d2}^{\text{real}}- L_{B1}- L_{B3})  }{-1+q_3}  -L_{B1}, \\
  	 \theta_1=&\sqrt{\frac{- 3 R_{56}^{s_0\to s_f}  }{3(1-  q_3)L_{d1}+(2- 3q_3   ) L_{B1}- q_3^2 L_{B3}  }}.
  	\end{aligned}
  \end{equation}

For asymmetric S-chicane:

After considering the dipole lengths,  the variables $ \ell_2= L_{d2}/L_{d1}, \ell_3= L_{d3}/L_{d1}$  can give new expressions   as
 \begin{equation}\label{eq:ell23}
\begin{aligned}
 	\ell_2=&\frac{4  }{q_3} +3( \frac{q_2}{q_3}+1)+ \frac{L_{B1}}{2 L_{d1}q_3}(7+6q_2+6q_3) \\
+& \frac{L_{B2}}{2 L_{d1}q_3}(4+3q_2+2q_3 )-  \frac{L_{B3}}{2 L_{d1} }-\frac{L_{B4}}{2 L_{d1}q_3} (  1+q_2+q_3 ) ,\\
\ell_3=& - \frac{4 }{q_3}-\frac{3q_2}{q_3}  - \frac{L_{B1}}{2 L_{d1}q_3}( 7+6q_2 ) - \frac{L_{B2}}{2 L_{d1}q_3}( 4+3q_2  )\\
-&  \frac{L_{B3}}{2 L_{d1} }+\frac{L_{B4}}{2 L_{d1}q_3} (  1+q_2+q_3 ) .   
\end{aligned}
 	\end{equation}
Here the added condition  $L_{d1}/L_{tot}=1/4$ is rewritten as $(L_{B1}+0.5L_{B2}+L_{d1})/L_{tot}=1/4$ in order to fix the position of the 2nd dipole when the   total chicane lengths are constant. Equation \eqref{eq:condition-position} is tenable if the $\ell_2, \ell_3$ in Eq. \eqref{eq:ell23} are degenerated by setting $L_{Bi}=0$. 
 One can obtain $L_{d1}$ and the bending angle $\theta_1$ from the value of $	R_{56}^{s_0\to s_f}$, $L_{tot}$  accroding to 
 \begin{equation}
	\begin{aligned}
		L_{tot}& =  4	(L_{B1}+0.5L_{B2}+L_{d1}) =4L_{B1}+2L_{B2}+4L_{d1},  \\
		R_{56}^{s_0\to s_f}&= \theta_1^2  \left( \left(   L_{d1}   q_2 +  L_{d3}   q_3  q_4 \right) + \frac{L_{B1}}{6}   ( 1+3q_2) + \frac{L_{B2}}{6}  q_2(3 + q_2)  \right)  \\
		&  + \frac{\theta_1^2 L_{B4} }{6}  ( 1+2  q_2+q_2^2-q_3-q_2q_3-2q_3^2 )\\
&-   \frac{\theta_1^2 L_{B3} }{6} q_3( 3  +3q_2+2 q_3).
	\end{aligned}	
\end{equation}

 	\section{Derivation of  the  Enteries of   R-matrix }\label{appendix1}
 	Considering the variation of bunch length within each dipole, the bunch parameters $k_1,k_2,k_3,k_4$ in Eq. \eqref{eq:deltai} are  no longer set to a fixed value for each dipole, but are modified as follows 
 
 \begin{equation}
 	k_i=\frac{k_0}{ (1+h~ R_{56}^{s_0\to s}) ^{4/3}},~~(i=1,2,3,4) 
 \end{equation}
 where  $ R_{56}^{s_0 \to s}$ in different  positions, as a function of $\phi$,   are expressed in Appendix \ref{appendix1}. Here  $\phi $ and the   subscript ``$s$" are the angle and position that the beam traverses in a
 dipole magnet, respectively.
 Thus  the CSR cancelation conditions  for a chicane can  be evaluated  using an integration method as    \cite{Emma:1997hj} 
 \begin{equation} \label{eq:integrate}
 	\begin{aligned}
 \Delta x_f^2=	\left[	\sum_{i=1 }^{ 4} \rho_i^{1/3} \int_{B} k_i  ~R_{16}^{s \rightarrow s_f}    d \phi \right]^2 =0, \\
  \Delta x_f^{\prime 2} =	\left[ 		\sum_{i=1 }^{ 4} \rho_i^{1/3} \int_{B} k_i  ~R_{26}^{s \rightarrow s_f}    d \phi  \right]^2 =0 ,
 	\end{aligned}
 \end{equation}
% \begin{equation}\label{eq:delta}
 %	\begin{aligned}
% 		\left\langle\Delta x_i^{2}\right\rangle&=\left(\rho_i^{1/3} \int_{B} k_i  ~R_{16}^{s \rightarrow s_f}    d \phi \right)^{2}, ~~~
% 		\left\langle\Delta x_i^{\prime 2}\right\rangle=\left(\rho_i^{1/3} \int_{B} k_i  ~R_{26}^{s \rightarrow s_f}    d \phi \right)^{2},\\
% 		&\left\langle\Delta x_i \Delta x_i^{\prime} \right\rangle=\rho_i^{2/3} \left( \int_{B} k_i  ~R_{16}^{s \rightarrow s_f}    d \phi \right)\left( \int_{B} k_i  ~R_{26}^{s \rightarrow s_f}    d \phi \right) .
% 	\end{aligned}
% \end{equation}
here  both    $R_{16}^{s \rightarrow s_f}$ and $R_{26}^{s \rightarrow s_f}$  vary  as a function of  $\phi $  in  dipoles.

 For	asymmetric C-chicane:

 	For $s$ within the 1st bend:

  	\begin{equation}\label{eq:r161}
\begin{aligned}
 	R_{16}^{s\to s_f}&= L_{d1}(\theta_1-\theta_{1s} )- (2L_{B3}+L_{d3}+L_{d2}) \theta_{1s}\\
&
+L_{B1} (\theta_1-2\theta_{1s} +\frac{\theta_{1s}^2}{2\theta_1}  ) + (L_{B2}+L_{d3})\theta_3 ,\\
R_{26}^{s\to s_f}&= - \theta_{1s } ,\\
R_{56}^{s_0 \to s}&=\frac{ L_{B1} \theta_{1s}^3}{ 6\theta_1}  ,
\end{aligned}
\end{equation}
 	with $\theta_{1s}=\theta_1 s/L_{B1}  $, and $0\le s\le L_{B1}$. 
 	
 	For $s$ within the 2nd bend:

  	\begin{equation}\label{eq:r162}
\begin{aligned}
 	R_{16}^{s\to s_f}&= - (\theta_1+\theta_{2s })(2L_{B3}+L_{d3}+L_{d2})- \frac{L_{B1}  (\theta_1+\theta_{2s})^2 }{2\theta_1} \\
& + (L_{B2}+L_{d3})\theta_3  ,\\
R_{26}^{s\to s_f}&= - (\theta_1+\theta_{2s })    ,\\
 	R_{56}^{s_0\to s}&= L_{d1} \theta_1 \theta_{2s } +  \frac{L_{B1}}{6} \left(\theta_1^2 + 3\theta_1  \theta_{2s}-3  \theta_{2s}^2 -\frac{\theta_{2s}^3 }{\theta_1  } \right),
\end{aligned}
\end{equation}
 	with $\theta_{2s}=-\theta_1 s/L_{B1}  $, and $0\le s\le L_{B1}$.

 For $s$ within the 3rd bend:

  	\begin{equation}\label{eq:r163}
\begin{aligned}
 R_{16}^{s\to s_f}&= L_{d3} (\theta_{3}-\theta_{3s }) + L_{B3}  \left( \theta_3  -2\theta_{3s}+\frac{\theta_{3s}^2}{2 \theta_3  }    \right),\\
 R_{26}^{s\to s_f}&= -\theta_{3s},\\
 R_{56}^{s_0\to s}&=   \frac{ L_{B3}\theta_{3s}^3  }{6\theta_{3}} + L_{d1} \theta_1 (-\theta_1+\theta_{3s})+L_{B1} (-\frac{2\theta_1^2}{3}+\theta_1 \theta_{3s}) ,
\end{aligned}
\end{equation}
 with $\theta_{3s}= \theta_3 s/L_{B3}  $, and $0\le s\le L_{B3}$. 
 
 For $s$ within the last bend:
  
  	\begin{equation}\label{eq:r164}
\begin{aligned}
&R_{16}^{s\to s_f}=  -\frac{ L_{B3}(\theta_3+ \theta_{4s})^2}{ 2\theta_3},\\
&R_{26}^{s\to s_f}=-\theta_3- \theta_{4s},\\
 &R_{56}^{s_0\to s}=L_{d1}\theta_1(-\theta_1+\theta_3+\theta_{4s}) +  L_{d3}\theta_3\theta_{4s}\\
&+  \frac{ L_{B1}\theta_1 }{3} \left( -2\theta_1+3\theta_3+3\theta_{4s}  \right) + \frac{ L_{B3}}{6} (\theta_3^2+3\theta_3\theta_{4s}-3\theta_{4s}^2- \frac{ \theta_{4s}^3 }{ \theta_{3}}      )      ,
\end{aligned}
\end{equation}
 with $\theta_{4s}= -\theta_3 s/L_{B4}  $, and $0\le s\le L_{B4}$.

 For	asymmetric S-chicane:

 	 For $s$ within the 1st bend:
 	 
 	 \begin{equation}
\begin{aligned}
 	  R_{16}^{s\to s_f}&=\frac{\theta_{1s}}{2   (\theta_1+\theta_2)}  \left(    -(L_{B2}+2L_{d1})\theta_2 +(L_{B3}+2L_{d3}) \theta_3  \right)  \\
 &	+\frac{ L_{B1} \theta_{1s}}{2 \theta_1 (\theta_1+\theta_2)}   \left(   -\theta_1^2+\theta_1(\theta_{1s}-2\theta_2)+\theta_{1s}\theta_2    \right) \\
&-\frac{\theta_{1s}}{2   (\theta_1+\theta_2)}L_{B4}(2\theta_1+ \theta_4),\\
 	 R_{26}^{s\to s_f}&= - \theta_{1s } ,\\
	 R_{56}^{s_0\to s}&=\frac{ L_{B1} \theta_{1s}^3}{ 6\theta_1} ,
 	 \end{aligned}
 \end{equation}  
 	 with $\theta_{1s}=\theta_1 s/L_{B1}  $, and $0\le s\le L_{B1}$. 
 	 
 	 For $s$ within the 2nd bend:
 	 
 \begin{equation}
\begin{aligned}
 	 R_{16}^{s\to s_f}&=\frac{1}{2}\left( 2(L_{B3}+L_{B4}+L_{d2}+L_{d3})(\theta_2-\theta_{2s})\right)\\
& +\frac{1}{2}\left( L_{B3}\theta_3+2 (L_{B4}+L_{d3})\theta_3+L_{B4} \theta_4+\frac{L_{B2}(\theta_2-\theta_{2s})^2}{\theta_2} \right),\\
 R_{26}^{s\to s_f}&= -\theta_1-\theta_{2s},\\
 R_{56}^{s_0\to s}&= L_{d1} \theta_1 \theta_{2s } +  \frac{L_{B1}}{6} \left(\theta_1^2 + 3\theta_1  \theta_{2s}  \right)+\frac{\theta_{2s}^2L_{B2}(3\theta_1+\theta_{2s})}{6\theta_2},
\end{aligned}
 \end{equation}
 with $\theta_{2s}= \theta_2 s/L_{B2}  $, and $0\le s\le L_{B2}$. 
 	 
 	 For $s$ within the 3rd bend:
  
 \begin{equation}
\begin{aligned}
R_{16}^{s\to s_f}&=  \frac{1}{2}    \left( 2(L_{B4}+L_{d3}) (\theta_3-\theta_{3s})+\frac{L_{B3}(\theta_3-\theta_{3s})^2}{  \theta_3  }+L_{B4} \theta_4   \right) , \\
 R_{26}^{s\to s_f}&=\theta_3+\theta_4-\theta_{3s},\\
 	 R_{56}^{s_0\to s}&=  \frac{1}{6 } \left(L_{B1}\theta_1^2+L_{B2}  \theta_2^2  +3\theta_1\theta_2(L_{B1}+L_{B2}+2L_{d1}) \right) \\
&  - \frac{\theta_{3s} }{2} \left(
 	     -\theta_4 (L_{B4}+2L_{d3})+L_{B3}(2 \theta_1+2 \theta_2 +\theta_3)  
 	 \right) \\
 	& + \frac{  L_{B3}\theta_{3s}^2 }{6\theta_3}  (3\theta_1+3\theta_2+\theta_{3s}) ,
 \end{aligned}
\end{equation}
 	 with $\theta_{3s}= \theta_3 s/L_{B3}  $, and $0\le s\le L_{B3}$. 
 	 
 	 For $s$ within the last bend:

\begin{equation}
\begin{aligned}
R_{16}^{s\to s_f}&= \frac{ L_{B4}(\theta_4- \theta_{4s})^2}{ 2\theta_4},\\
 R_{26}^{s\to s_f}&= \theta_4- \theta_{4s},\\
 	 R_{56}^{s_0\to s}&= - \frac{\theta_3}{6 } \left(     3L_{B3}(\theta_1+\theta_2)+2L_{B3}\theta_3-6(L_{B4}+L_{d3})\theta_4 +3L_{B4}\theta_4  \right)  \\
 	 &+\frac{1}{6 }(6L_{d1}\theta_1\theta_2 +L_{B2}\theta_2 (3\theta_1+\theta_2) +L_{B1}\theta_1 (\theta_1+3\theta_2)\\
&+ \frac{1}{6 }( 3L_{B4}\theta_4\theta_{4s}-3L_{B4}\theta_{4}\theta_{4s}^2  ) +\frac{1}{6\theta_4}   L_{B4}\theta_{4s}^3  ,
 \end{aligned}
\end{equation}
 	 with $\theta_{4s}= \theta_4 s/L_{B4}  $, and $0\le s\le L_{B4}$.

 %	\begin{figure}
 	%	\centering
 %		\includegraphics[scale=0.35]{bunch-length} 
 %		\caption{The bunch length variation  for a $C=10$ chicane. The  gray areas in all plots represent bending magnets and the area width represent the dipoles length.
 %		}
 %		\label{fig:bunch-length} 
 %	\end{figure} 

 \end{appendices}

\end{document}